\documentclass[11pt]{article}

\usepackage{graphicx} 
\usepackage[margin=1in]{geometry} 

\usepackage{comment}
\usepackage{bbm}
\usepackage[dvipsnames, table]{xcolor}
\usepackage{mathtools} 
\usepackage{natbib}
\usepackage{graphicx} 
\usepackage{listings}
\usepackage{enumerate}
\usepackage{scalerel} 
\usepackage{amsmath}
\usepackage{amssymb}
\usepackage{amsfonts}
\usepackage{booktabs}
\usepackage{multirow}
\usepackage{threeparttable}
\usepackage{subcaption}

\usepackage{float}
\usepackage{caption}
\usepackage{colortbl}
\newcommand{\nothere}[1]{}

\usepackage{orcidlink}

\newtheorem{thm}{Theorem}[section]

\title{Nonparametric estimation of the Patient Weighted While-Alive Estimand}
\author{Alessandra Ragni$^{1,*}$\orcidlink{0000-0002-3647-7340}, Torben Martinussen$^{2,**}$\orcidlink{0000-0002-9760-6791} and Thomas Scheike$^{2,***}$\orcidlink{0000-0002-2148-4740}        
\\
\small{$^{1}$MOX, Department of Mathematics, Politecnico di Milano} \\
\small{$^{2}$Section of Biostatistics, University of Copenhagen, Copenhagen, Denmark}      \\
\small{$^{*}$alessandra.ragni@polimi.it, $^{**}$tma@sund.ku.dk, $^{***}$thsc@sund.ku.dk}}
\date{}
\begin{document}

\maketitle

\abstract{
In clinical trials with recurrent events, such as repeated hospitalizations terminating with death, it is important to consider the patient events overall history for a thorough assessment of treatment effects. The occurrence of fewer events due to early deaths can lead to misinterpretation, emphasizing the importance of a while-alive strategy as suggested in Schmidli et al. (2023).
In this study, we focus on the patient weighted while-alive estimand, represented as the expected number of events divided by the time alive within a target window, and develop efficient estimation for this estimand.
Specifically, we derive the corresponding  efficient influence function and develop a one-step estimator initially applied to the simpler irreversible illness-death model. For the broader context of recurrent events, due to the increased complexity, this one-step estimator is practically intractable due to likely misspecification of the needed conditional transition intensities that depend on a patient's unique history.  Therefore, we suggest an alternative estimator that is expected to have high efficiency, focusing on the randomized treatment setting. 
Additionally, we apply our proposed estimator to two real-world case studies, demonstrating the practical applicability of this second estimator and benefits of this while-alive approach over currently available alternatives.
}

\vspace{0.3cm}


\section{Introduction}
\label{sec:introduction}

Recurrent events, such as repeated hospitalizations or episodes of a chronic condition, commonly occur in clinical studies and significantly impact patient outcomes and overall health trajectories.
In clinical trials and randomized experiments, it is important to consider the entire history of patient events to accurately assess clinical treatment effects.
The European Medicines Agency (EMA) in a qualification opinion emphasized that treatments are expected to impact not only the first event, but also subsequent ones, advocating for clinically meaningful measures of treatment effect based on recurrent event endpoints, allowing more insightful statistical analyzes compared to those focusing solely on the first event \citep{akacha2018request}.
Many statistical methods have been proposed for analyzing recurrent event data, such as \cite{prentice1981regression, andersen1982cox, lin1989robust, wei1989regression, lin2000semiparametric, liu2004shared, mao2016semiparametric}.
However, the development of estimands for recurrent events with a causal clinical interpretation has not yet been thoroughly explored  \citep{imbens2015causal, lipkovich2020causal}. For clarity, we refer to causal estimands as those defined within the potential outcomes framework, which requires envisioning the outcomes for a patient if assigned to the test treatment versus the outcomes if assigned to the control treatment \citep{imbens2015causal, pearl2016causal}.


Recently, \cite{roger2019treatment} proposed an estimator for the
treatment-policy estimand for recurrent event data, and
\cite{schmidli2023estimands}, building on the EMA's request, presented an
overview of different estimands for recurrent events terminated by death. Among
these, the while-alive (or while-on-treatment) strategy was proposed. 
While-alive estimands which are the focus of this paper aim to examine 
the treatment effect while patients are alive or, in other words, while intercurrent events —such as
treatment discontinuation, death, intake of rescue medication, or change of
background medication— have not occurred \citep{schmidli2023estimands,mao2023nonparametric}.
Specifically, the aim is to consider the time alive  when 
aiming to compare the number of recurrent events.
When death occurs, the time during which patients can experience recurrent events is shortened,
making the rate of these events more clinically  meaningful than the total
count. Consequently, in a clinical trial, if we consider the extreme case where
most patients die immediately under the control treatment while no patients die
under the test treatment, almost no events will be observed for the control
group but potentially many for the test treatment group. This discrepancy can
lead to misinterpretation of the treatment effect, highlighting the need for a
while-alive strategy.

While the average number of events observed in the presence of death as a semi-competing risk has earned much attention \citep{gray1988class, cook1997marginal, ghosh2000nonparametric, schaubel2010estimating, mao2016semiparametric, cortese2022efficient, baer2023causal, rytgaard2024nonparametric},
few developments have been proposed in the literature regarding the while-alive strategy.
\cite{wei2023properties}, in a paper related to the EMA request, approached the while-alive estimand proposed by \cite{schmidli2023estimands} mainly under parametric assumptions. Specifically, they derived the analytical expression for the while-alive event rate using a gamma frailty model. 
\nothere{Moreover, they explored various estimators, including quasi-Poisson regression, negative binomial regression, the Lin-Wei-Yang-Ying model with a minimal death rate, and the method-of-moments estimator, assuming uniform censoring time across all patients.}
Instead, \cite{mao2023nonparametric} developed a general nonparametric estimator for the \textit{Exposure-Weighted While-Alive} (EWWA) estimand within a robust inference framework. In particular, 
the focus is on a general class of while-alive estimands, which measure the instantaneous loss incurred by the incident (i.e., new) events with weights possibly dependent on past experience averged over the Restricted Mean Survival Time (RMST) within a target time window.
For the computation of a nonparametric estimator, techniques similar to those from the \cite{ghosh2000nonparametric} for the numerator and the Kaplan-Meier plug-in estimator for the RMST for the denominator are employed. 
By adjusting for exposure time using the RMST in the denominator, the EWWA estimand does not distinguish between patients, and, as a consequence,  may overlook individual-level relationship between events occurrence and survival time.

In this paper, we develop a semiparametrically efficient estimation for the \textit{Patient Weighted While-Alive} (PWWA) estimand, defined in \cite{schmidli2023estimands} as the expected number of events divided by the time alive up to a target time window.
To the best of our knowledge, this estimand has not been explored in detail before.
We first derive the corresponding efficient influence function (EIF) allowing us to develop  the one-step estimator \citep{van2000asymptotic, laan2003unified, kennedy2022semiparametric}
in a general form and discuss its robustness property. This estimator is semiparametrically efficient only if all required  working models are correctly specified. 
However, this is practically infeasible in the general recurrent events setting, given the need to implement and specify all required conditional transition intensities that depend on the patient's unique history.
This is only feasible in simpler cases, such as the illness-death setting.
Therefore, we alternatively propose a feasible efficient estimator focusing on the randomized treatment setting with a correctly specifiable
censoring pattern, such as the common administrative censoring.  In this setting, the
PWWA estimand can still be estimated consistently and the proposed estimator is guaranteed to have superior performance compared to the standard inverse probability weighted complete case estimator (IPWCC), see \cite{tsiatis2006semiparametric}.

The remainder of this paper is organized as follows.
In Section \ref{sec:setup_notation}, we define the PWWA estimand in a causal setting, recalling key concepts of recurrent events multi-state models. In Section \ref{sec:EIF}, we compute the efficient influence function  presenting the irreversible illness-death model as a subcase of the recurrent events setting. 
In Section \ref{sect:Estimation and inference}, after general considerations related to the estimation and inference, we propose a consistent and practically feasible estimator with high efficiency. In Section  \ref{sec:simstudy}, we set up the simulation study separately for the irreversible illness-death model, where the one-step estimator and the proposed estimator are compared, and recurrent events case, where results employing the proposed estimator are shown. The performance of the estimator for the PWWA estimand is compared to that of the EWWA estimand.
Furthermore, two applications to case studies related to patients with chronic heart failure and metastatic colorectal cancer are presented, respectively, in Section \ref{sec:case_study} and Web Appendix D.
Section \ref{sec:discussion} contains a discussion with some concluding remarks as well as possible future developments. Technical derivations are relegated to the Appendix and Web Appendices.
%
The proposed estimator is implemented in the \texttt{WA\_recurrent}-function, in the \texttt{mets} R package \citep{holst2016} and a demonstration version of the employed code is available at \href{https://github.com/alessandragni/PWWAestimand}{https://github.com/alessandragni/PWWAestimand}.

\section{The Patient Weighted While Alive Estimand}
\label{sec:setup_notation}

\begin{figure}[H]
    \centering
    \includegraphics[width=0.6\textwidth]{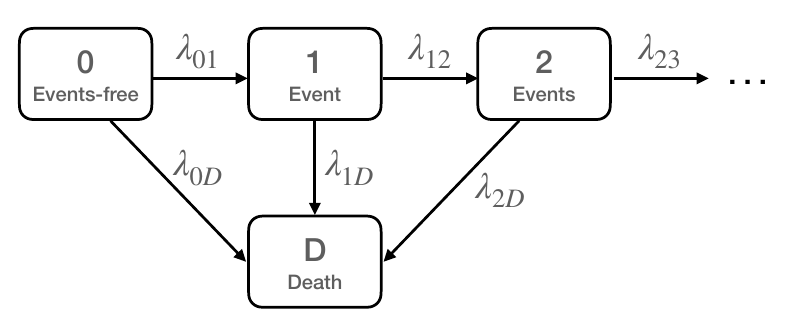}
    \caption{Recurrent events multi-state model with the $\lambda_{jk}$'s denoting the transition intensities between the different states.}
    \label{fig:Recurr}
\end{figure}

We consider a recurrent events multi-state model in a semi-competing risk setting, i.e., where non-terminal events compete with a terminal event \citep{fine2001semi,andersen2012statistical}.
Given a stochastic process $\{X(t)\}_{t\in[0, \tau]}$, $\tau<\infty$, with right-continuous sample paths, let $\{0,1,2,\dots,K,D\}$ be the finite state space, where $0$ may be considered as healthy state, $1,2,\dots,K$ states corresponding to recurrent non-terminal events (e.g., illnesses, relapses) and $D$ stands for the terminal event (death). We assume that $X(0) = 0$ and the only possible transitions are $0\rightarrow1$, $0\rightarrow D$, $1\rightarrow D$, $1\rightarrow 2$, $2\rightarrow D$, \dots, as depicted in Figure \ref{fig:Recurr}.

Let $T_1, T_2, \dots,T_K$ be the times to the non-terminal events and $T_D$ be the time to the terminal event since zero, respectively, and $\delta_k = \mathbbm{I}(T_k \leq T_D)$ for $k = 1, \dots, K$, where $\mathbbm{I}(\cdot)$ denotes the indicator function.
If the non-terminal event $k$ is not experienced before the terminal event, we define $T_k = +\infty$ (i.e., $\delta_k = 0$).
Employing the standard notation for counting processes, we define $N(t \wedge T_D) = \sum_{k=1}^{K} \mathbbm{I}(T_k \leq t, \delta_k = 1)$, which denotes the number of events before the terminal event in a target time window $[0,t]$ and where $a \wedge b = \min\{a,b\}$.

Let $A \in \{0,1\}$ denote the   treatment indicator and $L$ is a $p$-dimensional vector of baseline covariates.
We assume $X \sim P$, where $X=(A,L)$ and $P$ is a probability distribution belonging to a nonparametric statistical model $\mathcal{P}$.
The \textit{patient-weighted while-alive causal estimand }
may be expressed as
   $ \psi_t(P) = \mathbbm{E} ( Y_t^{a}) $, 
where 
$
Y_t=g \left \{ N(t \wedge T_D) / (T_D \wedge t) \right \}$.
Let $Y_t^{a}$ be the potential outcome of $Y_t$ had treatment been set to $a$.
Further, $g(\cdot)$ is a known function such as the identity function.
However, with this specific choice of $g$, as pointed out by  \cite{schmidli2023estimands}, the distribution of $\psi_t(P)$ is typically 
skewed due to early deaths. 
We present therefore the methodology with a general $g(\cdot)$. 
We assume that the set of covariates $L$ is sufficient for identification of the estimand $\psi_t(P)$  via the G-formula:
\begin{equation}
\label{eq:PWestimand}
    \psi_t(P) = \mathbbm{E} \{\mathbbm{E} ( Y_t | A = a, L)  \}.
\end{equation}
The causal contrast that we study is 
$
\mathbbm{E}( Y_t^{1}- Y_t^{0}),
$
which is referred to as an individual-level estimand because it provides a summarization of the causal effect defined at the individual level \citep{fay2024causal}. The EWWA estimand is based on $\mathbbm{E}\{N(t\wedge T_D)\}/\mathbbm{E}(T_D\wedge t)$ leading to a  contrast that cannot be written as a mean of two potential outcomes (treatment vs. control)  and is therefore not an individual-level causal estimand.

\section{Efficient influence function}
\label{sec:EIF}
We first give the full data EIF and then generalize it to the observed data case allowing for right-censoring.
Let $Z = \{T_K \wedge T_D, \delta_K, T_D, \overline{T}_{K-1}, X \}$ denote the full data meaning no censoring, with $\overline{T}_{K-1} = (T_1,\dots, T_{K-1})$ and $X=(A,L)$.
The EIF corresponding to the patient-weighted while-alive causal estimand  in  (\ref{eq:PWestimand}) is
\begin{align}
   D_{\psi}^*(P;Z) &=  \frac{\mathbbm{I}(A=a)}{\mathbbm{P}(A=a\mid L)}  \big\{Y_t - H_t(P;L) \big\}+ H_t(P;L) - \psi_t(P)\label{eif1Z}  \\
     &=     \omega(A,L)H_t(P;L) + b_t(P; Z) -\psi_t(P), \label{eif2Z} 
\end{align}
where  $H_t(P;L) =  \mathbbm{E} ( Y_t | A=a, L  )$, $\omega(A,L)= \big[ \mathbbm{P}(A=a\mid L)-\mathbbm{I}(A=a)\big] / \mathbbm{P}(A=a\mid L)$ and $b_t(P; Z)=Y_t \cdot \mathbbm{I}(A=a)/\mathbbm{P}(A=a\mid L)$.
It is clear from \eqref{eif1Z} that $\mathbbm{E}\{D_{\psi}^*(P_n;Z)\}=0 $ if we can correctly specify 
$H_t(P_n;L) $ while it is seen from \eqref{eif2Z} 
 that $\mathbbm{E}\{D_{\psi}^*(P_n;Z)\}=0 $ if we are able to correctly specify the propensity score model 
$P_n(A=a|L).$ Throughout, we use $P_n$ to indicate that working models have been applied to estimate  unknown quantities. 

We deal  now  with the observed data case allowing for right-censoring by $\widetilde C$. Let $\widetilde{T}_D = T_D \wedge \widetilde{C}$, $\delta_D = \mathbbm{I}(T_D \leq \widetilde{C})$,
$\widetilde{T}_k = T_k \wedge \widetilde{T}_D$ and $\delta_k = \mathbbm{I}(T_k \leq \widetilde{T}_D)$.
Then $O = \{ \widetilde{T}_D, \delta_D , \overline{\widetilde{T}}_{K}, \overline{\delta}_K, X \}
\sim P$ is the observed data where $P$ belongs to the non-parametric statistical model $\mathcal{P}$. Let $\widetilde N(t)= \sum_{k=1}^{K} \mathbbm{I}(\widetilde T_k \leq t, \delta_k = 1)$ denote the number of observed recurrent events at time point $t$.
We assume that data consist of $n$ iid replicates $O_1,\ldots ,O_n$.
The observed data is a result of monotone coarsening of the full data. To be specific,  we introduce the coarsening variable $\mathcal{C}$ such that when  $\mathcal{C}=r$ we only get to see a coarsened version $G_r(Z)$ of the full data. 
The observed data is thus equivalently expressed as 
$O = \{\mathcal{C}, G_\mathcal{C}(Z)\}$.
Let $K\{r \mid  G_{r}(Z)\} = \mathbbm{P}(\widetilde{C}>r \mid G_r(Z)) = \exp \{ - \int_0^r \lambda_{\widetilde{C}}\{s; G_s(Z)\} ds \}$ be the survival function 
corresponding to the censoring distribution, and $d M_{\widetilde{C}} \{r \mid G_r(Z)\} = dN_{\widetilde{C}}(r) - \mathbbm{I}(\widetilde{T}_D\geq r) d\Lambda_{\widetilde{C}}\{r \mid G_r(Z)\}$ be the increment of the censoring martingale,
where $\Lambda_{\widetilde{C}}\{r \mid G_r(Z)\} = \int_0^r \lambda_{\widetilde{C}}\{s; G_s(Z)\} ds$ and $N_{\widetilde{C}}(r) = \mathbbm{I}(\widetilde{T}_D \leq r, \delta_D=0)$.
The observed data efficient influence function is then given by
\begin{eqnarray}
       D_{\psi}^*(P;O) & = & \frac{\delta_Db_t(P; Z)}{K\{T_D|G_{T_D}(Z)\}}-\psi_t(P)+ \omega(A,L)H_t(P;L) \nonumber \\
   && + \int \mathbbm{E}\{b_t(P; Z)|G_r(Z)\}\frac{dM_{\widetilde{C}}\{r|G_r(Z)\}}{K\{r|G_r(Z)\}}, \label{eif1O}
\end{eqnarray}
which follows using \cite{tsiatis2006semiparametric}, formula (10.76), see the Appendix 
for more details. 
Note that $G_r(Z)$ always contains $(A,L)$ so 
$\mathbbm{E}\{b_t(P; Z)|G_r(Z)\}=\mathbbm{E}\{Y_t|G_r(Z)\} \cdot \mathbbm{I}(A=a)/\mathbbm{P}(A=a\mid L)$.
Because of the structure of  the efficient influence function 
 \eqref{eif1O}, we see that  the one-step estimator \citep{kennedy2022semiparametric} is given by 
\begin{equation}
\label{os-estimator}
    \widehat \psi_t^{os}=\mathbbm{P}_n\widetilde D_{\psi}(P_n;O),
\end{equation}
where 
$ \widetilde D_{\psi}(P;O)=D_{\psi}^*(P;O) +\psi_t(P)$ and  
$\mathbbm{P}_n\{v(Z)\}=n^{-1}\sum_i v(Z_i)$ denotes the empirical measure.  
In the next subsection, we give the specific expression for $D_{\psi}^*(P;O)$ in the illness-death setting that is a special case of the recurrent events setting with a terminal event.

{\textit{Remarks}}
\begin{enumerate}[(i)]
    \item It is seen from \eqref{eif1O}  that we obtain  consistent estimation if the working models for the propensity score and censoring are correctly specified as then $\mathbbm{E}\{ D_{\psi}^*(P_n;O)\}=0$.
    \item 
    In the Appendix, 
    we give two equivalent expressions for $ D_{\psi}^*(P;O)$. From there it is seen, \eqref{eif2O}  and \eqref{eif3O}, that  we also get consistent estimation if the working model for  $\mathbbm{E}\{Y_t|G_r(Z)\}$ is correctly specified and either the model for censoring or    the propensity score are correctly specified.
    \item We will argue in a moment  that it is practically  challenging  to correctly specify a model for $\mathbbm{E}\{Y_t|G_r(Z)\}$ or to otherwise obtain consistent estimation of this quantity.
\end{enumerate}
\vspace{-0.4cm}

\subsection{Irreversible illness-death model}
\label{subsec:illnessdeath}
We consider here  the irreversible illness-death model that is  a special case of the recurrent events case. For this simpler case it is possible to derive an explicit form of the EIF \eqref{eif1O}, which shows all the terms needed to estimate in order to obtain the one-step estimator  \eqref{os-estimator}.
The state space is reduced  to $\{0,1,D\}$ and the full data is
$Z = \{T_1 \wedge T_D, \delta_1, T_D, X \}$.
The hazard and cumulative hazard functions for the illness-death model are defined respectively as 
$\lambda_{01}(t_1)$
and  $\Lambda_{01}(t_1) = \int_{0}^{t_1} \lambda_{01}(s) ds$; 
$\lambda_{0D}(t_D)$
and $\Lambda_{0D}(t_D) = \int_{0}^{t_D} \lambda_{0D}(s) ds$; 
$\lambda_{1D}(t_D\mid t_1)$
and $\Lambda_{1D}(t_D \mid t_1) = \int_{0}^{t_D} \lambda_{1D}(s \mid t_1) ds$.
The observed data are  $O = \{ \widetilde{T}_D = T_D \wedge \widetilde{C}, \delta_D = \mathbbm{I}(T_D \leq \widetilde{C}), \widetilde{T}_1 = T_1 \wedge \widetilde{T}_D, \delta_1 = \mathbbm{I}(T_1 \leq \widetilde{T}_D), X \}$.
For this special case, we have that the last term in \eqref{eif1O} has the explicit form (see Web Appendix A for details):
\begin{align}
     &\quad  \int_0^{\widetilde{T}_1\wedge t} \exp\{\Lambda_{\mathbf{\scalebox{1.2}{$\cdot$}}}(r)\} \Bigg[ \int_r^t \int_{t_1}^t \frac{1}{g(u)}f_{1D}(t_1, u \mid t_1)du\exp\{ -\Lambda_{\mathbf{\scalebox{1.2}{$\cdot$}}}(t_1)\} d\Lambda_{01}(t_1) \nonumber \\
    & \quad + \frac{1}{g(t)}\int_r^{\tau}\int_{t_1\vee t}^{\tau} f_{1D}(t_1, u \mid t_1) du \exp\{ -\Lambda_{\mathbf{\scalebox{1.2}{$\cdot$}}}(t_1)\} d\Lambda_{01}(t_1) \Bigg] \frac{dM_{\widetilde{C}} \{r \}}{K_{\widetilde{C}}\{r\}}  \nonumber \\
    & \quad + \delta_1 \mathbbm{I}(\widetilde{T}_1\leq t)\bigg[ \int_{\widetilde{T}_1}^{\widetilde{T}_D\wedge t}   \int_r^{\tau}
         \frac{1}{g(u \wedge t)}  f_{1D}(r, u \mid T_1) du \frac{dM_{\widetilde{C}} \{r \mid T_1\}}{K_{\widetilde{C}}\{r\mid T_1\}} \nonumber \\
    &\quad + \frac{\mathbbm{I}(t<\widetilde{T}_D)}{g(t)} \int_{ t}^{\widetilde{T}_D}  \int_r^{\tau}
           f_{1D}(r, u \mid T_1) du  
   \frac{dM_{\widetilde{C}} \{r \mid  T_1\}}{K_{\widetilde{C}}\{r\mid  T_1\}}  \bigg], 
   \label{eq:at_FINA}
\end{align}
where 
$\Lambda_{\mathbf{\scalebox{1.2}{$\cdot$}}}(\cdot) =\Lambda_{01}(\cdot)+\Lambda_{0D}(\cdot)$,
$
f_{1D}(r, u \mid s) = \exp\bigl \{-\int_r^{u} d\Lambda_{1D}(v \mid  s)\bigr \}\lambda_{1D}(u \mid  s)
$,
$a \vee b = \max\{a,b\} $ and any distributional quantity depending on $P$ needs to be conditioned on $A=a,L$, which  we have suppressed for notational convenience.
Moreover, the $H_t(P;L)$ may be rewritten as (see Web Appendix A for details)
$$
H_t(P;L) =  \int_0^t \int_{t_1}^{\infty}\frac{f_{1D}(t_1, u \mid t_1, A=a, L)du}{g(u\wedge t)}    
\exp \big\{ -\Lambda_{\mathbf{\scalebox{1.2}{$\cdot$}}}(t_1\mid A=a, L) \big\}
d\Lambda_{01}(t_1\mid A=a, L).
\label{Ht_FINAL}
$$
If we were able to correctly specify all the required working models, then the one-step estimator \eqref{os-estimator} is semiparametrically efficient, and its variance can be estimated consistently using the variance of the corresponding EIF. In Section \ref{sec:simstudy}, we examine the numerical performance of the one-step estimator by referring to the standard decomposition of the one-step estimator into the plug-in term (i.e., $\mathbbm{P}_n\{H_t(P_n;L)\}$) and debiasing term, which can be easily retrieved by decomposing in (\ref{eif1O}), the term $\omega(A,L)$ defined in (\ref{eif2Z}). 

It is also clear from \eqref{eq:at_FINA} that this specific quantity is challenging to model correctly because it involves the cumulative hazard functions $\Lambda_{01}, \Lambda_{0D}$ and $\Lambda_{1D}$ with the latter conditional on $(A,L)$ and $T_1$.

\section{Estimation and inference in the recurrent events setting when treatment is randomized}
\label{sect:Estimation and inference}

As shown in Section \ref{subsec:illnessdeath}, the fully efficient estimator  involves conditional hazard functions that depend on a history unique to each patient. This  becomes very complex in the  general recurrent events setting
making  computation  and practical implementation of the fully efficient estimator very challenging and often unfeasible.
To address this issue, we propose an alternative   strategy aiming for a feasible estimator that is still expected to have high efficiency.
As pointed out earlier, 
despite incorrect models are applied for the complicated component $\mathbbm{E}\{b_t|G_r(Z)\}$,
we still obtain a consistent estimator if the propensity score and the censoring model are correctly specified.

This is feasible in specific yet important cases in which these models can be accurately specified, such as in randomized studies with simple censoring mechanisms such as progressive type 1 censoring (administrative censoring), where censoring occurs solely because patients remain alive by the time the data are analyzed.
In such scenarios, 
we can estimate the proposed estimand consistently despite the complicated structure of  efficient  influence function.
Further, as we shall see, we are able to develop the corresponding influence function of the suggested estimator,  which facilitates computation of  standard errors. We emphasize that the same strategy can be applied in more general settings, see Remark (ii) below Theorem \ref{thm:asympt.est}.
The key to the development of the proposed estimator is the following representation:
$       D_{\psi}^*(P;O)   = \bigl\{\omega(A,L)H_t(P;L) + b_t(P; Z)-\psi_t(P) \bigr\}
 -
  \int \left[b_t(P; Z)-\mathbbm{E}\{b_t(P; Z)|G_r(Z)\}\right] dM_{\widetilde{C}}\{r|G_r(Z)\}/K\{r|G_r(Z)\},
$
where the first term on the right-hand side of the latter equation (curly brackets) and the censoring martingale term are orthogonal. If we replace the unknown and involved terms $H_t(P;L)$ and $\mathbbm{E}\{b_t(P; Z)|G_r(Z)\}$ by linear functions we can the estimate these  so that the variance of the two terms in the EIF 
are  minimized separately. In what follows we restrict to the situation with a randomized treatment, which leaves the EIF unchanged as it lies in the (reduced) tangent space and thus still is the EIF. 
Now, define a $q$-dimensional time-dependent covariate vector $W_r$ containing $(A,L)$ but also $\widetilde N(r-)$  or some (other) known function of $\widetilde N(r-)$ and let $J_r=\mathbb{I}(r\leq \widetilde T_D)$ denote  the at risk indicator.
 We then replace 
   $\mathbbm{E}\{b_t(P; Z)|G_r(Z)\}$ with $\gamma^T(r)W_r$ where $\gamma(r)$ denotes a $q$-dimensional  time-dependent coefficient, and we also replace 
   $H_t(P;L)$ with $\theta^TL$. We then choose these regression coefficients so that the variance of the two terms in the EIF is minimial. We further assume that the censoring is independent as in the progressive type 1 censoring case.
   We also use the logistic model for the propensity score $\mathbbm{P}(A=1|L)$ even though it is known by design as this procedure results in an estimator with smaller variance.
We show in the Appendix 
and Web Appendix B that this leads to the following  estimator:
 \begin{align}
\label{prop.est}
   \widehat \psi_t=  \widetilde \psi_t
   +  \mathbbm{P}_n\left [\omega_n(A,L)\theta_n^TL+
 \int \gamma_n^T(r)W_r
  \frac{dM_{\widetilde{C}}^n\{r\}}{K_n\{r\}}\right ],
\end{align}
\noindent where 
\begin{equation}
     \widetilde \psi_t=\mathbbm{P}_n\left [\frac{\delta_Db_t(P_n; Z)}{K_n\{T_D\}}\right ],
     \label{eq:dr-estimator}
\end{equation}
 \begin{align*}
\gamma_n(r)&=\left\{\mathbbm{P}_n J_r(W_r-\overline{W}_r)(W_r-\overline{W}_r)^T\right\}^{-1} \mathbbm{P}_n\left\{\frac{J_r\delta_Db_t(P_n;Z)(W_r-\overline{W}_r)}{K_n\{T_D\}}\right\},\\
 \theta_n&=-\left [\mathbbm{P}_n\{\omega_n^2(A,L)LL^T\}\right ]^{-1}\mathbbm{P}_n\left\{\frac{\delta_D
b_t(P_n;Z)\omega_n(A,L)L}{K_n\{T_D\}}\right\}
\end{align*}
with $\overline{W}_r=\{\mathbbm{P}_nJ_rW_r\}/\{\mathbbm{P}_nJ_r\}.$
\noindent
As we use the  Kaplan-Meier estimator  for $K_n$, the second  term on the right hand side of \eqref{prop.est} can be written as
$$
\mathbbm{P}_n\int \gamma_n^T(r)\{W_r-\overline{W}_r\}
  \frac{dM_{\widetilde{C}}\{r\}}{K_n\{r\}}=\mathbbm{P}_n\int \gamma_n^T(r)\{W_r-\overline{W}_r\}
  \frac{dN_{\widetilde{C}}\{r\}}{K_n\{r\}}.
$$

\begin{thm}\label{thm:asympt.est}
 Consider
 i.i.d. replicates of  $O = \{ \widetilde{T}_D, \delta_D , \overline{\widetilde{T}}_{K}, \overline{\delta}_K, X \}$ 
 such that the treatment is randomized, and $K\{r|G_r(Z)\}=K(r)$.
 Then $n^{1/2}(\widehat \psi_t-\psi_t)=n^{-1/2}\sum_{i=1}^n\phi_{\psi}(P,O_i)+o_P(1)$ where the explicit expression of the influence function  $\phi_{\psi}(P,O)$ is 
  \begin{align*}
     \phi_{\psi}(P,O)=&\frac{\delta_Db_t(P; Z)}{K(T_D)}+\omega(A,L)\theta^TL-\psi_t+\{\mathbbm{E} D_{\alpha}V(\alpha)\}\phi_{\alpha}(A,L)\nonumber\\
     &+\int \left\{\gamma^T(r)(W_r-\overline{w}_r)+
      \mathbb{E}(b_t(P;Z)|T_D\geq r) \right \}
\frac{dM_{\widetilde{C}}(r)}{K(r)}
 \end{align*}
 Thus,  $n^{1/2}(\widehat \psi_t-\psi_t)$
 converges in distribution to a  normal distribution  with zero-mean 
and a variance that can be consistently estimated  by $\mathbbm{P}_n\phi_{\psi}(P_n,O)^2$.

 \mbox{ }\hfill $\Box$
\end{thm}

\noindent
The proof is given in the Web Appendix B.

\medskip

\noindent
{\textit{Remark}}

\begin{itemize}
    \item[(i)]
The proposed estimator $\widehat \psi_t=\widehat \psi_t\{\theta_n,\gamma_n(\cdot)\}$ 
is guaranteed to be more efficient than the IWPCC estimator $\widetilde \psi_t$ as the latter results when $\theta=0$ and $\gamma(\cdot)=0$, while the proposed $\theta_n$ and $\gamma_n(\cdot)$ are chosen so that the variance of the estimator
$ \widehat \psi_t\{\theta,\gamma(\cdot)\}$ is minimized; see the proof of Theorem \ref{thm:asympt.est}. We further show in Web Appendix B
that estimating the
propensity score model, even when known, leads to improved efficiency. 
\item[(ii)] In this section, we have restricted to the randomized treatment setting and also assumed simple random censoring but the same strategy can be applied in more general scenarios
as long as both the propensity score and censoring models can be correctly specified. 
We have outlined this in the Web Appendix B.
\end{itemize}

\section{Simulation Studies}
\label{sec:simstudy}

In this section, we demonstrate the numerical performances of the estimators described above. 
In Web Appendix C, we consider the irreversible illness-death model setting and demonstrate the robustness properties of the one-step estimator. We also make a comparison to the estimator proposed in Section
\ref{sect:Estimation and inference}. The simulation results show that the 
one-step estimator is more efficient;
however, there are no major differences between the estimators in the considered setting.

We next consider the case of recurrent events, where we adopt the consistent estimator 
with high efficiency  as described in (\ref{prop.est}).
We first generate data 
with recurrent events and the terminal event that mimics
that of the HF-Action trial \citep{OConnor2011}, which we return to in Section \ref{sec:case_study}.
Specifically, we use the observed rate of hospitalization (the recurrent events) and the rate of 
terminal event (death) from fitting Andersen-Gill rate models adjusting for treatment. The study had
$4$ years of follow up. 
The rate of hospitalization was roughly speaking approximately constant ($0.78$ per year) and the
rate for terminal event was approximately $0.07$ per year.  These are  the baseline rates in our data generating model taking using  rates:
%
$Z \lambda_1(t) \exp(A \beta_1+L \beta_L)$ for the recurrent events process, 
and  $Z^v \lambda_d(t) \exp(A \beta_d+L \beta_{d,L})$ for time to death.
Here, $A$ and $L$ are binary covariates drawn independently from Bernoulli distribution with success probability 0.5. 
Coefficients are set to $\beta_1=\beta_d = -0.3$ and $\beta_L = \beta_{d,L} = 0.3$, and $Z$, the frailty, 
is a gamma-distributed random variable with mean 1 and variance $\theta$. The power $v$ is $1$ or $0$ to generate dependence between 
$D$ and the recurrent events or not, respectively. When $v=1$ the frailty introduces dependence between the 
recurrent events and death. A large variance introduces high dependence and when the variance is $0$ then $Z\equiv 1$ and the two 
data-generating processes are independent. 
Finally, 
the censoring time is given by $\widetilde{C} \sim \text{Exp}(\lambda_{\widetilde{C}})$ 
with $\lambda_{\widetilde{C}} = k_{\widetilde{C}}/4$.

The results, using
$g(\cdot) = \sqrt[3]{\cdot}$, are evaluated at $t=3$, denoting the number of years 
since the start of the observation at which the response is evaluated. 
We consider both independent events and scenarios with shared random effects with
$\theta=\{0.5,1,2\}$. Additionally, different censoring rates
$k_{\widetilde{C}}=\{1,2\}$ are analyzed. 
The obtained results for the case where $v=1$ are summarized in
Table \ref{tab:case_recurrent_censored_estimatorALLCASES_SECOV}, obtained through 
\texttt{WA\_recurrent()} in the \texttt{mets} R package \citep{holst2016}, see also \href{https://github.com/alessandragni/PWWAestimand}{https://github.com/alessandragni/PWWAestimand}. 
The estimator in the table was derived using two augmentation model specifications: (i) a model that included $\widetilde{N}(r-)$, $L$, and $Z$ for censoring augmentation and both $L$ and $Z$ for the mean ratio model, and (ii) a simpler model that included only $\widetilde{N}(r-)$ and $L$ for censoring augmentation and only $L$ for the mean ratio model.
The estimation procedure was replicated 5000 times and the sample size is set to 1000.

\begin{table}
\centering 
\small
\begin{tabular}{@{}lccccccccccccccc@{}}
\toprule 
  & & & \multicolumn{4}{c}{$\widehat \psi_t$ in (\ref{prop.est}) -- (i) with $Z$} & \multicolumn{4}{c}{$\widehat \psi_t$ in (\ref{prop.est}) -- (ii) without $Z$} & \multicolumn{4}{c}{$\widetilde \psi_t$ in (\ref{eq:dr-estimator})}\\
\cmidrule(l{0pt}r{6pt}){4-7} \cmidrule(l{6pt}r{6pt}){8-11} \cmidrule(l{6pt}r{0pt}){12-15} 
$\theta$ & $k_{\widetilde{C}}$ & $A$ & \textbf{Mean} & \textbf{SD} & \textbf{SE} & \textbf{Cov} & \textbf{Mean} & \textbf{SD} & \textbf{SE} & \textbf{Cov}  & \textbf{Mean} & \textbf{SD} & \textbf{SE} & \textbf{Cov}\\ 
\midrule
\multirow{6}{*}{$0.5$} 
& \multirow{3}{*}{$1$} 
&   $0$ &  0.770 & 0.023 & 0.023 & 0.941 & 0.772 & 0.024 & 0.025 & 0.955 & 0.776 & 0.028 & 0.029 & 0.953 \\ 
& & $1$ &  0.663 & 0.023 & 0.023 & 0.945 & 0.666 & 0.025 & 0.025 & 0.945 & 0.669 & 0.029 & 0.029 & 0.948 \\ 
\cmidrule{3-3} 
& & $0$-$1$ &  0.106 & 0.031 & 0.031 & 0.950 & 0.106 & 0.035 & 0.035 & 0.953 & 0.107 & 0.040 & 0.040 & 0.952 \\
\cmidrule{2-15}
&  \multirow{3}{*}{$2$} 
&   $0$ &  0.756 & 0.029 & 0.030 & 0.911 & 0.763 & 0.031 & 0.032 & 0.941 & 0.776 & 0.040 & 0.040 & 0.946 \\ 
& & $1$ &  0.651 & 0.030 & 0.030 & 0.912 & 0.657 & 0.032 & 0.032 & 0.936 & 0.669 & 0.041 & 0.040 & 0.943 \\ 
\cmidrule{3-3} 
& & $0$-$1$ &  0.106 & 0.041 & 0.042 & 0.949 & 0.106 & 0.044 & 0.045 & 0.947 & 0.107 & 0.058 & 0.057 & 0.945 \\ 
\cmidrule{1-15}
\multirow{6}{*}{$1$} 
& \multirow{3}{*}{$1$} 
&   $0$ &  0.702 & 0.025 & 0.025 & 0.943 & 0.704 & 0.028 & 0.028 & 0.943 & 0.709 & 0.032 & 0.032 & 0.945 \\ 
& & $1$ &  0.605 & 0.025 & 0.024 & 0.938 & 0.607 & 0.027 & 0.027 & 0.944 & 0.611 & 0.032 & 0.031 & 0.946 \\ 
\cmidrule{3-3} 
& & $0$-$1$ &  0.097 & 0.032 & 0.032 & 0.945 & 0.097 & 0.039 & 0.039 & 0.950 & 0.098 & 0.046 & 0.045 & 0.944 \\  
\cmidrule{2-15}
&  \multirow{3}{*}{$2$} 
&   $0$ &  0.688 & 0.033 & 0.032 & 0.914 & 0.696 & 0.035 & 0.035 & 0.938 & 0.708 & 0.046 & 0.045 & 0.951 \\
& & $1$ &  0.593 & 0.032 & 0.032 & 0.908 & 0.600 & 0.035 & 0.034 & 0.927 & 0.611 & 0.045 & 0.044 & 0.942 \\
\cmidrule{3-3} 
& & $0$-$1$ &  0.095 & 0.044 & 0.044 & 0.951 & 0.096 & 0.049 & 0.049 & 0.948 & 0.096 & 0.064 & 0.064 & 0.944 \\ 
\cmidrule{1-15}
\multirow{6}{*}{$2$} 
& \multirow{3}{*}{$1$} 
&   $0$ & 0.604 & 0.026 & 0.027 & 0.945 & 0.607 & 0.030 & 0.030 & 0.949 & 0.611 & 0.035 & 0.036 & 0.951 \\
& & $1$ & 0.522 & 0.026 & 0.025 & 0.947 & 0.524 & 0.029 & 0.029 & 0.943 & 0.528 & 0.033 & 0.034 & 0.948 \\
\cmidrule{3-3} 
& & $0$-$1$ &0.081 & 0.032 & 0.033 & 0.957 & 0.083 & 0.042 & 0.042 & 0.950 & 0.083 & 0.048 & 0.049 & 0.952 \\ 
\cmidrule{2-15}
&  \multirow{3}{*}{$2$} 
&   $0$ & 0.590 & 0.034 & 0.034 & 0.911 & 0.598 & 0.038 & 0.038 & 0.938 & 0.611 & 0.049 & 0.050 & 0.950 \\
& & $1$ & 0.509 & 0.033 & 0.033 & 0.906 & 0.516 & 0.036 & 0.036 & 0.931 & 0.528 & 0.048 & 0.048 & 0.943 \\
\cmidrule{3-3} 
& & $0$-$1$ & 0.081 & 0.044 & 0.045 & 0.955 & 0.082 & 0.051 & 0.052 & 0.951 & 0.083 & 0.068 & 0.069 & 0.947 \\ 
\bottomrule
\end{tabular}
\caption{Simulation results for the \textit{consistent estimator with high efficiency} $\widehat{\psi}_t$ in the recurrent events setting with $g(\cdot) = \sqrt[3]{\cdot}$, evaluated across different values of $\theta$ and $k_{\widetilde{C}}$. Two augmentation model specifications are considered: (i) including $Z$, and (ii) excluding $Z$. For $\widehat{\psi}_t$, we report the mean across simulations (Mean), standard deviation (SD), empirical standard error (SE), and coverage at 95\% confidence level (Cov). Results for its component $\widetilde{\psi}_t$ (from Eq.~\ref{eq:dr-estimator}) are shown for comparison. Each scenario is based on 5000 replications with a sample size of 1000.
}
\label{tab:case_recurrent_censored_estimatorALLCASES_SECOV}
\end{table}

We see that $\widehat \psi_t$ has smaller standard errors than $\widetilde \psi_t$, reflecting efficiency gain from the augmentation. Furthermore, incorporating $Z$ into the augmentation models leads to additional reductions in standard errors, indicating further efficiency gain by accounting for the heterogeneity. 
Across all scenarios, empirical standard deviations closely matched the average standard errors with coverage rates near the nominal 0.95 level.
The standard errors increase with increasing $\theta$ and censoring rate, with a larger increase in $\widetilde \psi_t$ than in $\widehat \psi_t$.




\subsection{Dependence}
\label{sec:dependence}


We now contrast the PWWA approach 
and the EWWA approach.
Specifically, we focus on how different types of dependence between $D$ and the recurrent events
affect the estimands. 
Indeed, one key challenge in this context is that early deaths result in shorter
risk periods for observing events when the recurrent events and the terminal event are
negatively correlated (their underlying hazards being positively correlated), which
is often the case in practical settings. If we focus on situations with constant
rates, the PWWA estimand, which normalizes $N(t\wedge T_D)$ by $T_D \wedge t$ at
the individual level, effectively accounts for this challenge, highlighting more
differences as the dependence between events and survival time increases, in
contrast to the EWWA estimand, 
that is the ratio of exposure-weighted event rates, $\mathbbm{E} \{N(T_D \wedge t)\} / \mathbbm{E} \{T_D \wedge t\}$.


We first considered a scenario corresponding to the HF-Action trial setting, as used in the previous simulations.
We refer to this as scenario (a).
We then introduced two alternative scenarios: (b) the baseline hazard for the recurrent events is piecewise constant with values 0.5 for $t\leq 1$ 
and 0.89 for $1<t\leq 4.31$, and (c) the baseline hazard for the recurrent events is piecewise constant with values 2.5 for $t\leq 1$ and 0.29 for $1<t\leq 4.31$. 
%
The hazard for the terminal event is kept the same as in the HF-Action setting. 
In all scenarios, we varied the event rates using the scaling factor
$s_D = \{1, 4\}$ for the terminal event. 
%
As described above, we consider two different types of dependence in the data: $v=1$ 
and $v=0$. 
%
In both cases, we augment with $\widetilde{N}(r-)$, $L$ and $Z$. 
We focus on $t=3$, $k_{\widetilde C} = 1$, $\theta = \{1,2\}$, $\beta_1 = \beta_d = -0.3$ and $\beta_L = \beta_{d,L} = 0.3$, replicating the estimation procedure 5000
times. Results for the causal contrast  (0-1) are shown in Table
\ref{tab:simulationstudyTab2}.


\begin{table}
\centering 
\small
\begin{tabular}{@{}llcccccccccc@{}}
\toprule 
& & & & \multicolumn{4}{c}{PWWA} & \multicolumn{4}{c}{EWWA} \\ 
\cmidrule(l{3pt}r{3pt}){5-8} \cmidrule(l{3pt}r{3pt}){9-12}
& $v$ & $\theta$ & $s_D$  & \textbf{Mean} & \textbf{SD} & \textbf{SE}  & \textbf{Power} & \textbf{Mean} & \textbf{SD} & \textbf{SE} & \textbf{Power}\\ 
\midrule
\multirow{8}{*}{(a)} & \multirow{4}{*}{$1$} 
& \multirow{2}{*}{1} 
   &  1 & 0.098 & 0.032 & 0.032 & 0.864 & 0.204 & 0.068 & 0.067 & 0.864 \\ 
& & & 4 & 0.084 & 0.035 & 0.035 & 0.680 & 0.152 & 0.060 & 0.059 & 0.737 \\ 
\cmidrule{3-12}
& & \multirow{2}{*}{2} 
  & 1   & 0.082 & 0.032 & 0.033 & 0.706 & 0.178 & 0.077 & 0.077 & 0.637 \\
& & & 4 & 0.070 & 0.034 & 0.034 & 0.538 & 0.111 & 0.060 & 0.059 & 0.457 \\
\cmidrule{2-12}
& \multirow{4}{*}{$0$} 
& \multirow{2}{*}{1} 
  & 1   & 0.096 & 0.032 & 0.032 & 0.854 & 0.241 & 0.077 & 0.074 & 0.900 \\ 
& & & 4 & 0.079 & 0.034 & 0.033 & 0.661 & 0.239 & 0.081 & 0.080 & 0.849 \\ 
\cmidrule{3-12}
& & \multirow{2}{*}{2} 
  & 1   & 0.080 & 0.032 & 0.033 & 0.688 & 0.240 & 0.097 & 0.096 & 0.709 \\ 
& & & 4 & 0.066 & 0.033 & 0.033 & 0.505 & 0.239 & 0.103 & 0.102 & 0.646 \\ 
\midrule
\multirow{8}{*}{(b)} & \multirow{4}{*}{$1$} 
& \multirow{2}{*}{1} 
   &  1 & 0.091 & 0.033 & 0.033 & 0.801 & 0.187 & 0.066 & 0.065 & 0.819 \\
& & & 4 & 0.070 & 0.036 & 0.035 & 0.512 & 0.124 & 0.055 & 0.055 & 0.619 \\
\cmidrule{3-12}
& & \multirow{2}{*}{2} 
  & 1   & 0.075 & 0.033 & 0.033 & 0.621 & 0.160 & 0.073 & 0.075 & 0.574 \\ 
& & & 4 & 0.059 & 0.035 & 0.034 & 0.410 & 0.089 & 0.055 & 0.054 & 0.382 \\ 
\cmidrule{2-12}
& \multirow{4}{*}{$0$} 
& \multirow{2}{*}{1} 
  & 1   & 0.090 & 0.032 & 0.032 & 0.801 & 0.226 & 0.074 & 0.073 & 0.873 \\ 
& & & 4 & 0.064 & 0.034 & 0.033 & 0.487 & 0.203 & 0.075 & 0.077 & 0.761 \\ 
\cmidrule{3-12}
& & \multirow{2}{*}{2} 
  & 1   & 0.075 & 0.032 & 0.033 & 0.628 & 0.227 & 0.095 & 0.094 & 0.676 \\ 
& & & 4 & 0.054 & 0.032 & 0.033 & 0.366 & 0.204 & 0.100 & 0.099 & 0.549 \\ 
\midrule
\multirow{8}{*}{(c)} & \multirow{4}{*}{$1$} 
& \multirow{2}{*}{1} 
   &  1 & 0.122 & 0.031 & 0.032 & 0.975 & 0.318 & 0.089 & 0.087 & 0.957 \\
& & & 4 & 0.135 & 0.034 & 0.035 & 0.973 & 0.325 & 0.097 & 0.096 & 0.926 \\
\cmidrule{3-12}
& & \multirow{2}{*}{2} 
  & 1   & 0.105 & 0.033 & 0.034 & 0.888 & 0.296 & 0.107 & 0.108 & 0.790 \\ 
& & & 4 & 0.115 & 0.035 & 0.036 & 0.904 & 0.262 & 0.105 & 0.105 & 0.713 \\ 
\cmidrule{2-12}
& \multirow{4}{*}{$0$} 
& \multirow{2}{*}{1} 
  & 1   & 0.118 & 0.031 & 0.032 & 0.963 & 0.347 & 0.092 & 0.090 & 0.973 \\ 
& & & 4 & 0.132 & 0.035 & 0.035 & 0.968 & 0.446 & 0.111 & 0.110 & 0.986 \\ 
\cmidrule{3-12}
& & \multirow{2}{*}{2} 
  & 1   & 0.100 & 0.034 & 0.035 & 0.826 & 0.348 & 0.116 & 0.117 & 0.853 \\ 
& & & 4 & 0.111 & 0.036 & 0.037 & 0.867 & 0.448 & 0.141 & 0.142 & 0.895 \\ 
\bottomrule
\end{tabular}
\caption{Results comparing PWWA and EWWA estimands across different simulation settings (a–b-c), dependence structures, and scaling factors, setting $\beta_1 = \beta_d = -0.3$. For each estimand, we report the average estimated causal effect for the contrast (0-1) (Mean), the standard deviation (SD), the empirical standard error (SE) and observed power, 
computed testing the null hypothesis of no causal contrast between groups, using a significance level of 0.05.
The sample size is set to 1000 and the estimation procedure is replicated 5000 times. 
}
\label{tab:simulationstudyTab2}
\end{table}

\noindent
Generally, when $v=0$, that is, the frailty only affects the intensity of the recurrent events, the EWWA is more powerful than the PWWA. On the other hand, when
 the frailty is fully shared across all event intensities ($v=1$) then the PWWA is generally more powerful than the EWWA especially with increasing 
 value of the variance $\theta$ of the frailty variable. 
Also, an early high recurrent event rate (scenario (c)) favors the PWWA compared to the EWWA in the case where there is dependence between the two processes ($v=1$).
Generally, increasing the rate of death means a drop in power for both procedures, except 
in scenario (c) with $v=1$ and $\theta=2$, where the PWWA has a (slight) increase in power  while the power still drops for the EWWA.


We also examine the null 
case $\beta_1 = \beta_d = 0$, assuming no causal effect of the treatment; 
results are reported in Web Table 3. 
In all scenarios, both estimators have a power at the nominal level of 0.05, as expected.





\section{HF-Action randomized controlled trial}
\label{sec:case_study}

\begin{figure}
         \centering
         \includegraphics[width=0.7\textwidth]{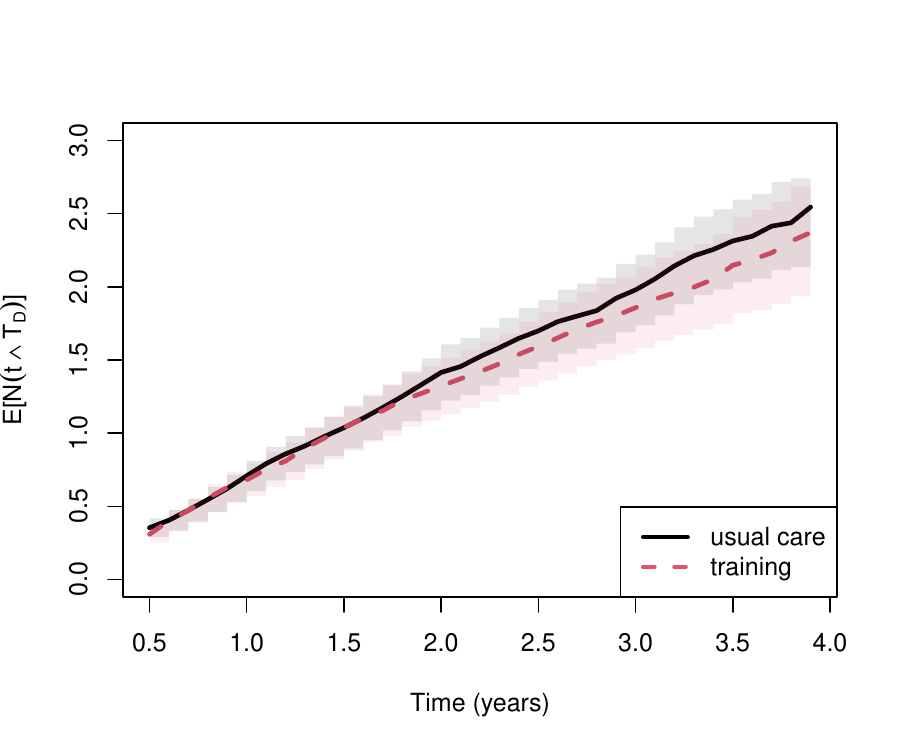}
     \caption{\textit{HF-Action randomized controlled trial}. Graphical visualization of $\mathbbm{E}\{ N(t\wedge T_D)\}$ (and $95\%$ confidence intervals) over 7 semesters.}
     \label{fig:hf-Action-NUM}
\end{figure}

\begin{figure}
     \centering
     \begin{subfigure}[b]{0.48\textwidth}
         \centering
         \includegraphics[width=\textwidth]{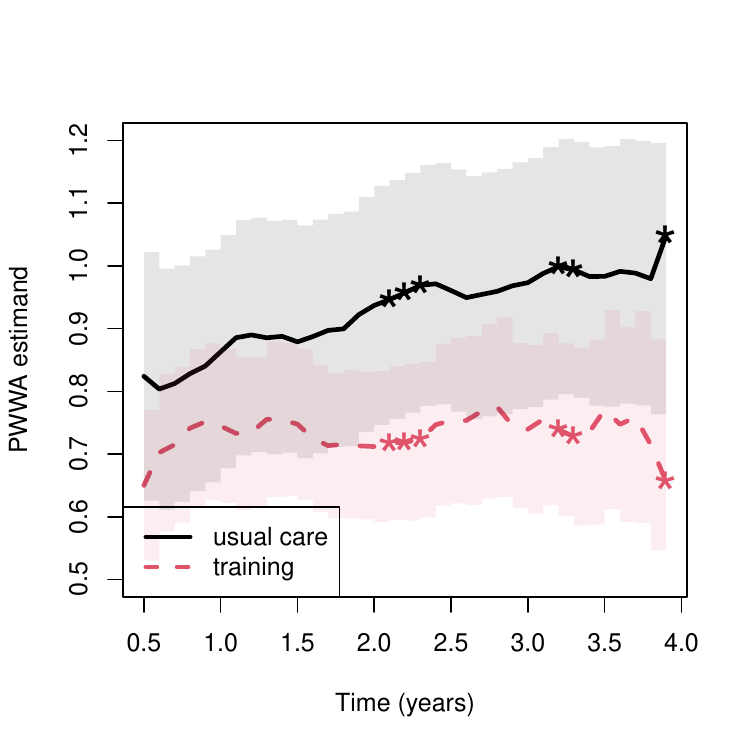}
         \caption{PWWA estimand.}
     \end{subfigure}
     \hfill
     \begin{subfigure}[b]{0.48\textwidth}
         \centering
         \includegraphics[width=\textwidth]{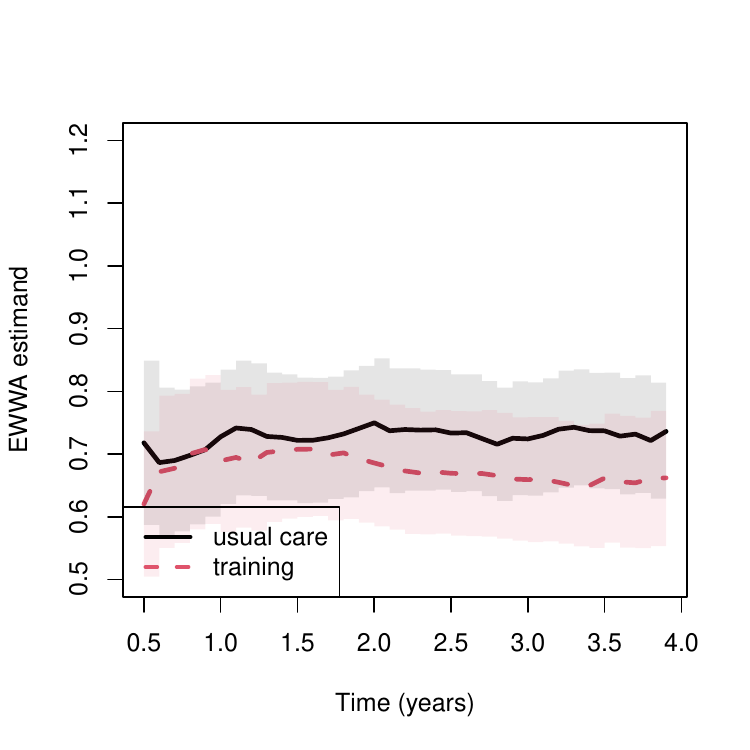}
         \caption{EWWA estimand.}
     \end{subfigure}
     \vspace{0.25cm}
     \caption{\textit{HF-Action randomized controlled trial}. Graphical visualization of PWWA and EWWA estimands (and their $95\%$ confidence intervals) over 7 semesters. Asterisks indicate a p-value $< 0.05$ for the test with null hypothesis \lq\lq no difference among usual care and training\rq\rq.}
     \label{fig:hf-Action}
\end{figure}

We illustrate our proposal using data from the 
HF-Action randomized controlled trial 
\citep{OConnor2011}, 
a study aimed at  investigating the effect of exercise training (treatment) compared to usual care in patients affected by chronic heart failure (HF) due to systolic dysfunction.
More specifically, the objective was to determine whether aerobic-type exercise training reduces 
all-cause hospitalization and improves quality of life.
The study was conducted as a multicenter covariate stratified block-randomized
study, with strata given by centers and heart failure etiology (ischemic vs
non-ischemic).
We here consider only those patients with the ischemic etiology (500 patients in each arm), 
and augmented our estimator using  the covariates: sex, center, age, beck depression (score), and previous heart
hospitalizations in 6 months prior to study (at baseline). 
The estimated mean number of hospitalizations along with 95\% confidence intervals are shown in Figure \ref{fig:hf-Action-NUM} for the two treatment arms that do not show a significant effect of
exercise training.
However, survival improved in the treatment group (HR $0.72$, 95\% CI $(0.52,1.01)$), making the interpretation of the mean number of hospitalizations subtle, since a prolonged life span creates more opportunities for recurrent hospitalizations.
This motivates us to use instead the PWWA estimand. For this application, we took $g(\cdot)$ as the identity
and thus consider 
the effect of exercise training on the mean number of hospitalizations per year before the terminal event over the time window $[0,t]$ years, with $t=0.5, \dots, 3.9$. 
The estimate is  shown 
in Figure \ref{fig:hf-Action}, left panel.

\nothere{
\begin{table}
\centering 
\small
\begin{tabular}{@{}llcccccc@{}}
\toprule 
& & \multicolumn{2}{c}{$t=2.3$} & \multicolumn{2}{c}{$t=3.3$} &  \multicolumn{2}{c}{$t=3.9$} \\ 
\cmidrule(l{3pt}r{3pt}){3-4} \cmidrule(l{3pt}r{3pt}){5-6} \cmidrule(l{3pt}r{3pt}){7-8} 
& & Est (SE) &  $p$-value & Est (SE) & $p$-value & Est (SE) & $p$-value \\ 
\midrule
\multirow{3}{*}{$g(\cdot) = \cdot$} 
  &   usual care (0)  & 0.969 (0.098) & 0 & 0.994 (0.104) & 0 &  1.048 (0.116) & 0 \\
& exercise training  (1) & 0.723 (0.063) & 0 & 0.728 (0.072) & 0 & 0.656 (0.079) & 0 \\
\cmidrule{2-8}
& Difference (0 - 1) & 0.246 (0.115) & 0.032 & 0.266 (0.125) &  0.034  & 0.393 (0.138) & 0.005 \\
\bottomrule
\end{tabular}
\caption{\textit{HF-Action randomized controlled trial} results for the PWWA estimand, for $g(\cdot)=\cdot$. Est, estimate of the PWWA estimand; SE, standard error.
}
\label{tab:case_studyHF}
\end{table}
}

\noindent
We see from Figure   \ref{fig:hf-Action}, left panel, that the 
mean number of hospitalizations per year before the terminal event is roughly constant over time but at a lower level in the exercise training arm than in the control arm.
At some of the time-points 
there is sufficient evidence to reject the null hypothesis, allowing us to conclude that treatment strategies have a statistically significant impact, and exercise training should be preferred.

We also estimated the EWWA, shown in
Figure  \ref{fig:hf-Action}, right panel, which 
reveals notable differences between the two estimands in this study.
The EWWA, which compares the average events with restricted mean survival for those in training versus those receiving usual care, shows no significant difference 
(Table \ref{tab:comparison_case_studyHF})
However, when employing the PWWA, a significant difference emerges, particularly for time points beyond two years.  This highlights the potential long-term benefits of the training versus the usual care, that can be better captured by an individual-level estimand such as the PWWA.

\begin{table}[H]
\centering 
\small
\begin{tabular}{@{}llcccccc@{}}
\toprule 
& & \multicolumn{2}{c}{$t=2.3$} & \multicolumn{2}{c}{$t=3.3$} &  \multicolumn{2}{c}{$t=3.9$} \\ 
\cmidrule(l{3pt}r{3pt}){3-4} \cmidrule(l{3pt}r{3pt}){5-6} \cmidrule(l{3pt}r{3pt}){7-8} 
& & Est (SE) &  $p$-val & Est (SE) & $p$-val & Est (SE) & $p$-val \\ 
\midrule
\multirow{3}{*}{Numerator} 
  &   usual care (0)  & 1.584 (0.104) & 0 & 2.212 (0.137) & 0 &  2.545 (0.164) & 0 \\
& training (1)   & 1.474 (0.109) & 0 & 1.998 (0.150) & 0 &  2.371 (0.200) & 0 \\
\cmidrule{2-8}
& Difference (0 - 1) &  0.110 (0.150) & 0.463 & 0.214 (0.203) & 0.291 & 0.175 (0.259)  & 0.500 \\
\midrule
\multirow{3}{*}{Denominator} 
  &   usual care (0)   & 2.146 (0.020) & 0 & 2.979 (0.035) & 0 & 3.458 (0.046) & 0 \\
& training (1)   &  2.201 (0.016) & 0 & 3.076 (0.030) & 0 & 3.581 (0.040) & 0 \\
\cmidrule{2-8}
& Difference (0 - 1) & -0.055 (0.026) & 0.033 &  -0.097 (0.046) & 0.036 & -0.124 (0.061)  & 0.042 \\
\midrule
Ratio (EWWA) & Difference (0 - 1) & 0.068 (0.070) & 0.329 & 0.093 (0.068) & 0.174 & 0.074 (0.075)  & 0.322 \\
\bottomrule
\end{tabular}
\caption{\textit{HF-Action randomized controlled trial}. Results for the EWWA estimand. Numerator, mean number of recurrent events up to time $t$ and is computed through Ghosh-Lin IPCW Cox-type model.
Denominator, RMST and is computed through IPCW regression.
Est, respective estimate; SE, standard error.
}
\label{tab:comparison_case_studyHF}
\end{table}

\section{Discussion}
\label{sec:discussion}

In this work, we investigated the patient-weighted while-alive estimand within a nonparametric framework, addressing the presence of right-censoring to reflect real-world scenarios where patients may drop out. We derived the 
efficient influence function 
and proposed two estimators.
The first estimator is fully efficient in theory. While its performance can be readily assessed through simulation studies in the specific case of the illness-death model, practical challenges arise in the broader context of recurrent events.
Indeed, the fully efficient estimator involves conditional hazard functions that depend
on a history unique to each patient, resulting in high risk of misspecification of the needed working models.
To mitigate this problem, we also proposed a feasible estimator.
While it allows for potential misspecification of the conditional hazard functions that depends on patient's history, it remains consistent as long as both the propensity score and the censoring model are correctly specified. Moreover, it is expected to achieve high efficiency and, importantly, is guaranteed to outperform the standard IPWCC estimator in terms of efficiency.
This estimator, applied to the recurrent events setting, demonstrates good performance in simulations.
Furthermore, two real-world randomized trials showcased the utility of the patient-weighted while-alive approach, highlighting its practical relevance in diverse clinical contexts when compared to other state-of-the-art methods.
Indeed, a key advantage of this method is its ability to account for the time each patient remains alive, instead of the average one, enabling to accurately capture its dynamic relationship with recurrent events and treatment effects.

It is of interest to study summary measures other than the mean of the patient-weighted-while-alive distribution, as this distribution is typically right-skewed. Such measures could include specific quantiles; however, this is left for future work.

\section*{Competing interests}
No competing interest is declared.

\section*{Data availability}
The HF-Action study data analyzed in Section \ref{sec:case_study} 
that support the findings of this study are available from the Biologic Specimen and Data Repository
Information Coordinating Center (BioLINCC) 
of the National Heart, Lung, and Blood Institute. Restrictions apply to the availability of these data, which were used under license for this study. 
The colorectal cancer dataset analyzed in Web Appendix D is instead available at 
\href{https://github.com/alessandragni/PWWAestimand}{https://github.com/alessandragni/PWWAestimand}. 

\section*{Online Supporting Information}
With this paper are also available:
Web Appendix A (referenced in Section~\ref{subsec:illnessdeath}), Web Appendix B (referenced in Section~\ref{sect:Estimation and inference}), Web Appendix C (referenced in Section \ref{sec:simstudy}),
Web Appendix D (referenced in Section \ref{sec:introduction}) and
Web Table 3 (referenced in Section \ref{sec:dependence}).
A demonstration version of the code used in the paper is available at \href{https://github.com/alessandragni/PWWAestimand}{https://github.com/alessandragni/PWWAestimand}. 

\section*{Acknowledgements}

The HF-Action trial data were kindly made available 
by the Biologic Specimen and Data Repository Information Coordinating Center (BioLINCC) 
of the National Heart, Lung, and Blood Institute.

\begin{appendix}
\counterwithin{equation}{section}
\counterwithin{figure}{section}
\counterwithin{table}{section}
\renewcommand\theequation{\thesection\arabic{equation}}

\section[Different representation of D]{Different representation of $D_{\psi}^*(P;0)$}
\label{eif_O}

The observed data efficient influence function can be expressed in different ways:
\begin{align}
       D_{\psi}^*(P;O)  & = \frac{\mathbbm{I}(A=a)}{\mathbbm{P}(A=a\mid L)}  \big\{Y_t - H_t(P;L) \big\}+ H_t(P;L) - \psi_t(P) \notag \\
  &\quad -
  \int \left[b_t(P; Z)-\mathbbm{E}\{b_t(P; Z)|G_r(Z)\}\right]\frac{dM_{\widetilde{C}}\{r|G_r(Z)\}}{K\{r|G_r(Z)\}}\label{eif2O}\\
  & = \omega(A,L)H_t(P;L) + b_t(P; Z)-\psi_t(P) \notag \\
  &\quad -
  \int \left[b_t(P; Z)-\mathbbm{E}\{b_t(P; Z)|G_r(Z)\}\right]\frac{dM_{\widetilde{C}}\{r|G_r(Z)\}}{K\{r|G_r(Z)\}},\label{eif3O}
\end{align}
The full data EIF can be written as in (\ref{eif2Z})
and it forms the basis for constructing estimating equations by inverse probability  weighting it and adding elements from the so-called augmentation space $\Gamma^{\mbox{A}}=\left\{h_t(r;G_r(Z)) : \int h_t(r;G_r(Z))dM_{\widetilde{C}}\{r|G_r(Z)\}/ K\{r|G_r(Z)\}\right \}$
(AIPWCC), see \cite{tsiatis2006semiparametric}: 
\begin{equation}
\label{aipwcc}
 \phi(P;O)=\frac{\delta_D D_{\psi}^*(P;Z)}{K\{T_D,G_{T_D}(Z)\}}+\int h_t(r;G_r(Z))\frac{dM_{\widetilde{C}}\{r|G_r(Z)\}}{K\{r|G_r(Z)\}}.   
\end{equation}
By varying the element in the augmentation space 
gives a class of influence functions defined by $D_{\psi}^*(P;Z)$ with the efficient one being
\begin{align*}
-\Pi\left (\frac{\delta_D D_{\psi}^*(P;Z)}{K\{T_D,G_{T_D}(Z)\}}\bigg | \Gamma^{\mbox{A}}\right )
\end{align*}
corresponding to $h_t(r;G_r(Z))
    =\mathbbm{E}\{D_{\psi}^*(P;Z) |G_r(Z)\}$
with \eqref{aipwcc} then giving
 $D_{\psi}^*(P;O)$. As we have argued in Section \ref{sect:Estimation and inference}, this optimal choice leads in general to an intractable estimator, however. 
 We will therefore seek an optimal solution in a restricted class of influence function in order to obtain a feasible estimator with high efficiency. In fact, it is guaranteed to have superior performance compared to the standard IPWCC estimator. 

\end{appendix}

\bibliographystyle{chicago} 
\bibliography{bibliography}

\end{document}


\maketitle

\renewcommand\thesection{\Alph{section}}

\section{Web Appendix A}
\subsection*{Computation of the EIF with right-censored data in the illness-death setting}

Following Chapters 7.1 and 9.3 in \cite{tsiatis2006semiparametric}, introduce a coarsening variable $\mathcal{C}$, i.e., a continuous random variable equal to the censoring time when $\widetilde{C}<T_1 \wedge T_D$ or $T_1<\widetilde{C}\leq T_D$, and equal to $\infty$ when the data is uncensored.
Let $\tau$ be a time horizon chosen such that there exists $\epsilon >0$ with $\mathbbm{P}(\widetilde{C} > \tau) >\epsilon > 0$. Then $\forall r \in [0, \tau]$, we define a many-to-one function of the full data
\begin{equation}
    G_r(Z) = 
    \begin{cases}
        (T_1 \wedge T_D \geq r, X) \quad & \text{if } (\widetilde{T}_1 = \widetilde{T}_D = \widetilde{C}, \delta_1 = 0, \delta_D = 0 )\\
        (\delta_1=1, T_1 < r \leq T_D, T_1, X) \quad & \text{if } (\widetilde{T}_1 = T_1, \widetilde{T}_D = \widetilde{C}, \delta_1 = 1, \delta_D = 0)\\
        (T_1 \wedge T_D, \delta_1, T_D, X ) \quad & \text{if full-data case } (r = \infty) 
        \label{eq:many-to-one}
    \end{cases}
\end{equation}
where the first case corresponds to \lq\lq censored before any event", while the second one to \lq\lq non-terminal event then censored prior to terminal event".
This leads to a situation of monotone coarsening since $G_r(Z) \subseteq G_{r'}(Z)$ for $r>r'$.
The observed data may now be expressed as $O = \{\mathcal{C}, G_\mathcal{C}(Z)\}$.

The full-data EIF $D_{\psi}^*(P;Z)$ may be mapped into the observed-data one $D_{\psi}^*(P;O)$ by the linear operator that transforms terms of the EIF affected by coarsening (because observed) and leaves unchanged terms that are functions of the full data, giving (by Theorems 10.1 and 10.4 in \cite{tsiatis2006semiparametric})
$$ \frac{\mathbbm{I}(A=a)}{\mathbbm{P}(A=a\mid L)} \cdot \big[a_t(P; O) - H_t(P;L) \big] + H_t(P;L) - \psi_t(P).  $$
where
    \begin{equation}
    a_t(P; O) = \frac{\delta_D \, a_t(P; Z)}{K_{\widetilde{C}}\{\widetilde{T}_D ; G_{\widetilde{T}_D}(Z)\}} + \int \frac{\mathbbm{E}\big[ a_t(P; Z) \mid G_{r}(Z) \big]}{K_{\widetilde{C}}\{r; G_{r}(Z)\}} dM_{\widetilde{C}} \{r ; G_r(Z)\}
    \label{eq:thm10.1&10.4}
    \end{equation}
being $K_{\widetilde{C}}\{r; G_{r}(Z)\} = \mathbbm{P}(\widetilde{C}>r \mid G_r(Z)) = \exp \{ - \int_0^r \lambda_{\widetilde{C}}\{s; G_s(Z)\} ds \}$ the conditional survival function 
and $d M_{\widetilde{C}} \{r ; G_r(Z)\} = dN_{\widetilde{C}}(r) - \mathbbm{I}(\widetilde{T}_D\geq r) d\Lambda_{\widetilde{C}}\{r; G_r(Z)\}$ the increment of the censoring martingale 
where $\Lambda_{\widetilde{C}}\{r; G_r(Z)\} = \int_0^r \lambda_{\widetilde{C}}\{s; G_s(Z)\} ds$ and $N_{\widetilde{C}}(r) = \mathbbm{I}(\widetilde{T}_D \leq r, \delta_D=0)$.
We aim to compute
    \begin{align}
    a_t(P; O) & = \frac{g(Y_t^{(1)}) \cdot \delta_D }{K_{\widetilde{C}}\{\widetilde{T}_D ; G_{\widetilde{T}_D}(Z)\}} + \int \mathbbm{E}\big[ g(Y_t^{(1)}) \big| G_{r}(Z) \big] \frac{dM_{\widetilde{C}} \{r ; G_r(Z)\}}{K_{\widetilde{C}}\{r; G_{r}(Z)\}} \nonumber \\
    & = \frac{g(Y_t^{(1))} \cdot \delta_D }{K_{\widetilde{C}}\{\widetilde{T}_D ; G_{\widetilde{T}_D}(Z)\}} + \int_0^{\widetilde{T}_1} \mathbbm{E}\big[ g(Y_t^{(1)}) \big| (T_1 \wedge T_D \geq r, X) \big] \nonumber \\
    & \quad \cdot \frac{dM_{\widetilde{C}} \{r ; (T_1 \wedge T_D \geq r, X)\}}{K_{\widetilde{C}}\{r;(T_1 \wedge T_D \geq r, X)\}} +\delta_1 \int_{\widetilde{T}_1}^{\widetilde{T}_D} \mathbbm{E}\big[ g(Y_t^{(1)}) \big| (\delta_1=1, T_1 < r \leq T_D, T_1, X) \big] \nonumber \\
    & \quad \cdot \frac{dM_{\widetilde{C}} \{r ; (\delta_1=1, T_1 < r \leq T_D, T_1, X)\}}{K_{\widetilde{C}}\{r;(\delta_1=1, T_1 < r \leq T_D, T_1, X)\}} \label{eq:a_t(O)}
    \end{align}
where the first equality is given by result in Eq.(\ref{eq:thm10.1&10.4}) and the second one is due to monotone coarsening in Eq. (\ref{eq:many-to-one}).
The expectations within the two integrals may be computed separately. 
For ease of notation, the conditioning with respect to $X$ will be omitted in the following. Moreover, we recall data are observed in the interval $[0, \tau]$, where $\tau$ refers to the end of the study.
For the first integral in $[0, \widetilde{T}_1]$, we get 
\begin{align*}
        & \mathbbm{E}\bigg[ \frac{\mathbbm{I}(T_1 \leq t, \delta_1 = 1)}{g(T_D \wedge t)} \big| (T_1 \wedge T_D \geq r) \bigg] \\ 
        & = \int \int \frac{\mathbbm{I}(t_1 \leq t, \delta_1 = 1)}{g(t_D \wedge t)} \frac{\mathbbm{P}(T_1 = t_1, T_D = t_D, T_1 \geq r, T_D \geq r)}{\mathbbm{P}(T_1 \geq r, T_D \geq r)} dt_1 dt_D \\
        & = \mathbbm{I}(r\leq t) \int_r^{\tau} \int_{r}^{t_D} \frac{\mathbbm{I}(t_1 \leq t, \delta_1 = 1)}{g(t_D \wedge t)} \frac{\mathbbm{P}(T_1 = t_1, T_D = t_D)}{\mathbbm{P}(T_1 \geq r, T_D \geq r)} dt_1 dt_D \\
        & = \mathbbm{I}(r\leq t) \int_r^{\tau} \frac{1}{g(t_D \wedge t)} \int_{r}^{t_D \wedge t} \frac{f(t_1, t_D)}{S(r,r)} dt_1 dt_D \\ 
        & = \mathbbm{I}(r\leq t) \int_r^{\tau} \frac{1}{g(t_D \wedge t)} \int_{r}^{t_D \wedge t}  \frac{S(t_1, t_1) \lambda_{01}(t_1) \lambda_{1D}(t_D\mid t_1) \exp \big\{ - \int_{t_1}^{t_D} \lambda_{1D}(u\mid t_1) du \big\} }{S(r,r)} dt_1 dt_D \\
        & = \frac{\mathbbm{I}(r\leq t)}{\exp \big\{ -\Lambda.(r) \big\} } \int_r^{\tau} \frac{1}{g(t_D \wedge t)} \int_{r}^{t_D \wedge t} \exp \big\{ -\Lambda.(t_1) \big\} \lambda_{01}(t_1) \lambda_{1D}(t_D\mid t_1) \\
        & \quad \cdot \exp \Big\{ - \int_{t_1}^{t_D} \lambda_{1D}(u\mid t_1) du \Big\}  dt_1 dt_D \\
        & = \frac{\mathbbm{I}(r\leq t)}{\exp \big\{ -\Lambda.(r) \big\} } \Bigg[ \int_r^t \int_{t_1}^t \frac{1}{g(t_D)} \exp \Big\{ - \int_{t_1}^{t_D} d\Lambda_{1D}(u\mid t_1)  \Big\} d\Lambda_{1D}(t_D\mid t_1) \\
        & \quad \cdot  \exp \big\{ -\Lambda.(t_1) \big\} d\Lambda_{01}(t_1) + \\
        & \quad + \frac{1}{g(t)}\int_r^{\tau}\int_{t_1\vee t}^{\tau} \exp \Big\{ - \int_{t_1}^{t_D} d\Lambda_{1D}(u\mid t_1) \Big\} d\Lambda_{1D}(t_D\mid t_1) \exp \big\{ -\Lambda.(t_1) \big\} d\Lambda_{01}(t_1) \Bigg] 
    \end{align*}
where for fourth equality we used Eq. (\ref{eq:joint-density}), in fifth equality we used Eq. (\ref{eq:joint-survival}), and in sixth equality we changed the order of integration (where $\max\{a,b\} = a \vee b$).
For the second integral in $[\widetilde{T}_1, \widetilde{T}_D]$, we get
    \begin{align*}
        & \mathbbm{E}\bigg[ \frac{\mathbbm{I}(T_1 \leq t, \delta_1 = 1)}{g(t_D \wedge t)} \big| (\delta_1=1, T_1 < r \leq T_D, T_1) \bigg] \\  
        & = \int \int 
         \frac{\mathbbm{I}(t_1 \leq t) \mathbbm{I}(t_1 \leq t_D)}{g(t_D \wedge t)} \mathbbm{P} ( T_1 = t_1, T_D = t_D \mid  T_1, T_1<r, T_D \geq r) dt_1 dt_D \\
         & = \mathbbm{I}(r<t) \int_r^{\tau}
         \frac{1}{g(t_D \wedge t)} \mathbbm{P} (T_D = t_D \mid  T_1, T_D \geq r) dt_D + \\
         & \quad + \mathbbm{I}(r\geq t) \mathbbm{I}(T_1 \leq t) \int_r^{\tau}
         \frac{1}{g(t)} \mathbbm{P} (T_D = t_D \mid  T_1, T_D \geq r) dt_D  \\
         & = \mathbbm{I}(r<t) \int_r^{\tau}
         \frac{1}{g(t_D \wedge t)} \exp \Big\{ - \int_r^{t_D} d\Lambda_{1D}(v \mid  T_1) \Big\} d\Lambda_{1D}(t_D \mid  T_1) + \\
         & \quad + \mathbbm{I}(r\geq t) \mathbbm{I}(T_1 \leq t) \int_r^{\tau}
         \frac{1}{g(t)} \exp \Big\{ - \int_r^{t_D} d\Lambda_{1D}(v \mid  T_1) \Big\} d\Lambda_{1D}(t_D \mid  T_1)
    \end{align*}
where for last equality we employed the following result
\begin{equation*}
        \mathbbm{P}(T_D = t_D \mid  T_1 = u, T_D > r) =\lambda_{1D}(t_D \mid  u) \text{ exp}\Big\{ - \int_r^{t_D} \lambda_{1D}(v \mid  u) dv \Big\} \, .
\end{equation*}
which follows from the fact that transition probabilities in an illness-death model are known and can be expressed in terms of the hazards of the transitions \cite{putter2007tutorial}.
Summing up these two terms within Eq. (\ref{eq:a_t(O)}), we get the desired result. 

\subsection*{Computation of $H_t(P;L)$ in the illness-death setting}
\label{app:comput_Ht}

Let $f(t_1, t_D)$ be the joint density of $T_1$ and $T_D$ in the upper wedge $0 < t_1 \leq t_D$, $f_{\infty}(t_D)$ the density of $T_D$ along $t_1 = + \infty$ for $t_D>0$, and $S(t_1, t_D)$ the bivariate survival function of $T_1$ and $T_D$ in the upper wedge \cite{Xu2010}.
Following \cite{Xu2010, lee2015bayesian, zhang2024marginal}, it can be proven that  
\begin{align}
    f(t_1, t_D) & = \lim_{\Delta \rightarrow 0} \lim_{\delta \rightarrow 0} \frac{\mathbbm{P}(T_1 \in [t_1, t_1 + \delta), T_D \in [t_D, t_D + \Delta))}{\Delta \cdot \delta} \nonumber \\
    & = S(t_1, t_1) \lambda_{01}(t_1) \lambda_{1D}(t_D \mid t_1) \, \exp \Big\{ - \int_{t_1}^{t_D} \lambda_{1D}(u\mid t_1) du \label{eq:joint-density}\Big\} 
\end{align}
where 
\begin{equation}
\label{eq:joint-survival}
    S(t,t) = \exp \big\{ -[\Lambda_{01}(t) + \Lambda_{0D}(t)] \big\} := \exp \big\{ -\Lambda.(t) \big\} \, .
\end{equation}

Employing these results, the expectation $H_t(P;L) = \mathbbm{E} \big[ g(Y_t^{(1)}) \big| A=a, L  \big]$ can be computed as follows:
\begin{align*}
        & \mathbbm{E}\bigg[ \frac{\mathbbm{I}(T_1 \leq t, \delta_1 = 1)}{g(T_D \wedge t)} \big| A=a, L \bigg] \\
        & = \int \int \frac{\mathbbm{I}(t_1 \leq t, \delta_1 = 1)}{g(t_D \wedge t)} f ( t_1, t_D \mid  A=a, L) dt_1 dt_D \\
        & = \int \int \frac{\mathbbm{I}(t_1 \leq t) \mathbbm{I}(t_1 \leq t_D)}{g(t_D \wedge t)} \exp \big\{ -\Lambda.(t_1 \mid A=a, L) \big\} \lambda_{01}(t_1 \mid A=a, L)  \\
        & \quad \cdot \lambda_{1D}(t_D \mid t_1,  A=a, L) \exp \Big\{ - \int_{t_1}^{t_D} \lambda_{1D}(u\mid t_1, A=a, L) du \Big\} dt_1 dt_D \\
        & = \int_0^t \int_{t_1}^{\infty} \frac{1}{g(t_D \wedge t)} \exp \Big\{ - \int_{t_1}^{t_D} \lambda_{1D}(u\mid t_1, A=a, L) du \Big\} \lambda_{1D}(t_D \mid t_1,  A=a, L) dt_D \\
        & \quad \cdot \exp \big\{ -\Lambda.(t_1 \mid A=a, L) \lambda_{01}(t_1 \mid A=a, L) \big\} dt_1 
    \end{align*}
where for the second equality we used Eq. (\ref{eq:joint-density}-\ref{eq:joint-survival}) and in third equality indicator functions were employing for setting up the extremes of integrations and terms have been reordered.

\newpage 

\section{Web Appendix B}
\subsection*{Proof of Theorem 4.1}
Let $\eta=(\theta^T,\gamma^T)^T$ suppressing the dependency of time in the notation, and denote the propensity score model by $\mathbbm{P}_n(A=1|L)=\pi(L)=e^{\alpha^TL}/(1+e^{\alpha^TL})$ (here letting $L$ include an intercept term)  with $\alpha_n$ denoting the MLE of $\alpha$.
The proposed estimator is $\widehat\psi_t=\psi_t(\widehat\eta)$, where we choose $\widehat \eta$ so that the variance of $\widehat \psi_t(\eta)$ is minimized at  $\widehat \eta$. For fixed $\eta$, we have 
$$
\widehat \psi_t(\eta)=\mathbbm{P}_n\left [\frac{\delta_Db_t(P_n; Z)}{K_n\{T_D\}}+ \omega_n(A,L)\theta^TL
+\int \gamma^T(r)W_r
  \frac{dM_{\widetilde{C}}^n\{r\}}{K_n\{r\}}\right ],
$$
which, following \citep{bang2000estimating}, can be rewritten as 
\begin{eqnarray*}
n^{1/2}\{\widehat \psi_t(\eta)- \psi_t\} & = &  n^{1/2}\mathbbm{P}_n\left [b_t(P_n; Z)+\omega_n(A,L)\theta^TL-\psi_t \right ]\\
&& +n^{1/2}\mathbbm{P}_n\Big[\int \Big\{ \gamma^T(r)(W_r-\overline{W}_r)-\Big(b_t(P_n;Z) \\
&& - \mathbbm{P}_n\frac{\mathbb{I}(\widetilde{T}_D\geq r)\delta_Db_t(P_n;Z)}{S_n(r)K_n(T_D)}\Big) \Big\}
\frac{dM_{\widetilde{C}}(r)}{K(r)}\Big]+o_p(1)\\
& =  &  n^{1/2}\mathbbm{P}_n\left [b_t(P; Z)+\omega(A,L)\theta^TL-\psi_t \right ]
+\{\mathbbm{E} D_{\alpha}V(\alpha)\}n^{1/2}\{\alpha_n-\alpha\}\\
&& +n^{1/2}\mathbbm{P}_n\Big[\int \big\{\gamma^T(r)(W_r-\overline{w}_r)-\big(b_t(P;Z) \\
&& - \mathbb{E}(b_t(P;Z)|T_D\geq r) \big)\big\}
\frac{dM_{\widetilde{C}}(r)}{K(r)}\Big]+o_p(1)\\
& = & B_1^n+B_2^n+B_3^n+o_p(1),
\end{eqnarray*}
where $\overline{W}_r=\{\mathbbm{P}_nJ_rW_r\}/\{\mathbbm{P}_nJ_r\}$, $J_r=\mathbb{I}(r\leq \widetilde T_D)$, $V(\alpha)=b_t(P_n; Z)+\omega_n(A,L)\theta^TL-\psi_t$ and $\overline{w}_r$ is the limit in probability of $\overline{W}_r$.
Furthermore, $n^{1/2}\{\alpha_n-\alpha\}=n^{-1/2}\sum_i\phi_{\alpha}(A_i,L_i)+o_p(1)$ with
$$
\phi_{\alpha}(A,L)=-E[\{A-\pi(L)\}^2LL^T]^{-1}L\{A-\pi(L)\}
$$
the influence function corresponding to the estimator $\alpha_n$. The two first terms on the right hand side of the latter display are independent of the third term. Also, asymptotically, $\mbox{var}(B_2^n)=-\mathbbm{E}(B_1^nB_2^n)$ so that $\mbox{var}(B_1^n+B_2^n)=\mbox{var}(B_1^n)-\mbox{var}(B_2^n)$, which shows that we get a more efficient estimator by estimating the propensity score.
Asymptotically, the variance of $n^{1/2}\{\widehat \psi_t(\eta)- \psi_t\}$ is $\mbox{var}(B_1^n)+\mbox{var}(B_3^n)-\mbox{var}(B_2^n)$ with the latter term not depending on $\eta$. Thus, the optimal $\theta$ is found by minimizing 
$$
E\{(b_t(P; Z)+\omega(A,L)\theta^TL-\psi_t)^2\}
$$
giving 
$$
\theta_t=\mathbbm{E}\{\omega(A,L)^2LL^T\}^{-1}\mathbbm{E}\{b_t(Z)\omega(A,L)L\}
$$
and we can further exploit that  
$$\mathbbm{E}\{b_t(Z)\omega(A,L)L\}=\mathbbm{E}\bigg\{\frac{\delta_Db_t(Z)\omega(A,L)L}{K(T_D)}\bigg\}.$$
 Using martingale calculus, one 
 similarly finds that  the optimal $\gamma_t(r)$ is the one  that solves
   $$
   0=\mathbbm{E}\left [\left \{
   \gamma^T(r)(W_r-\overline{W}_r)-\left (b_t(P;Z)
- \mathbb{E}(b_t(P;Z)|T_D\geq r)\right )\right \}(W_r-\overline{W}_r)^TJ_r
   \frac{d\Lambda_{\widetilde{C}}(r)}{K^2(r)}\right ],
   $$
with  $J_r=\mathbb{I}(r\leq \widetilde T_D)$  the at risk indicator. This leads to the optimal
 $$
{\gamma}_t(r)= \mathbbm{E}\left\{ (W_r-\overline{W}_r)(W_r-\overline{W}_r)^TJ_r\right\}^{-1}
\mathbbm{E}\left \{b_t(P; Z)(W_r-\overline{W}_r)J_r\right \}.
 $$
 Let $\widehat \eta$ denote this optimal $\eta$ and let $\widehat \psi_t=\psi_t(\widehat\eta)$.
It follows by simple calculations that $n^{1/2}\{\widehat \psi_t(\eta)- \psi_t\}$ and $n^{1/2}\{\widehat \psi_t- \psi_t\}$
 has the same limiting distribution because of the censoring and propensity score models being correctly specified.
 The influence function of $\widehat\psi_t$ is 
 \begin{align}
     \label{IF}
     \phi_{\psi}(P,O)=&\frac{\delta_Db_t(P; Z)}{K(T_D)}+\omega(A,L)\theta^TL-\psi_t+\{\mathbbm{E} D_{\alpha}V(\alpha)\}\phi_{\alpha}(A,L)\nonumber\\
     &+\int \left\{\gamma^T(r)(W_r-\overline{w}_r)+
      \mathbb{E}(b_t(P;Z)|T_D\geq r) \right \}
\frac{dM_{\widetilde{C}}(r)}{K(r)}
 \end{align}
 
 This concludes the proof.
 \mbox{ }\hfill $\Box$
\bigskip

\noindent
We now outline how to use the same estimation strategy as above in the setting where the treatment is not randomized and where censoring is different from simple random censoring. However, it requires that we correctly specify the propensity score model $\pi(L)$ and the censoring model. 
We consider a situation in which we have independent censoring, given $X=(A,L^T)^T$ and 
where the censoring hazard is given by $\lambda_{\widetilde C}(r|X)=X^T\beta(r)$. We use the Aalen least squares estimator $\widehat B(r)$, see \citep{martinussen2006dynamic}, to estimate the cumulative regression function $B(r)=\int_0^r\beta(s)\, ds.$
For fixed $\eta$, we have 
$$
\widehat \psi_t(\eta)=\mathbbm{P}_n\left [\frac{\delta_Db_t(P_n; Z)}{K_n\{T_D|X\}}+ \omega_n(A,L)\theta^TL
+\int \gamma^T(r)W_r
  \frac{dM_{\widetilde{C}}^n\{r|X\}}{K_n\{r|X\}}\right ],
$$
and we are seeking
 $\widehat \eta$ so that the variance of $\widehat \psi_t(\eta)$ is minimized at  $\widehat \eta$. The optimal $\theta$ turns out to be unchanged except that we need to use $K_n(T_D|X)$ instead of
 $K_n(T_D)$ in the final expression for the optimal  $\theta$ to take into account that the censoring depends on $X$. We now sketch how to find the optimal $\gamma(r)$. Let $Y_r$ be the matrix with $i$th row $J_{ir}X_i^T$. Using a Taylor expansion we then have
 \begin{align*}
     \frac{\delta_Db_t(P_n; Z)}{K_n\{T_D|X\}}=\frac{\delta_Db_t(P_n; Z)}{K\{T_D|X\}}+\frac{\delta_Db_t(P_n; Z)X^T}{K_n\{T_D|X\}}\int J_r\left (Y_r^TY_r\right )^{-1}Y_r^Td\vec{M}_{\widetilde C}(r|\vec{X})     
 \end{align*}
 plus a lower order term. We use the notation $\vec{M}_{\widetilde C}$ for the ($n\times 1$) vector consisting of the $n$ individual censoring martingale terms, and similarly with $\vec{X}$. Also,
 $$
 \frac{\delta_Db_t(P_n; Z)}{K\{T_D|X\}}=b_t(P_n; Z)-\int b_t(P_n; Z)\frac{dM_{\widetilde C}(r|X)}{K\{r|X\}}
$$ 
which gives 
 \begin{align*}
\mathbbm{P}_n  \frac{\delta_Db_t(P_n; Z)}{K_n\{T_D|X\}}=&\mathbbm{P}_n b_t(P_n; Z)\\
&-
\mathbbm{P}_n \int \left \{b_t(P_n; Z)-\mathbbm{P}_n\left (\frac{J_r\delta_Db_t(P_n; Z)X^T}{K_n\{T_D|X\}} \right )V_r^{-1}XK\{r|X\}
\right \}\frac{dM_{\widetilde C}(r|X)}{K\{r|X\}}
 \end{align*}
 with $V_r$ the limit in probability of $\mathbbm{P}_n(J_rXX^T).$
 We also have
 $$
\mathbbm{P}_n   \int \gamma^T(r)W_r
 \frac{dM_{\widetilde{C}}^n\{r|X\}}{K_n\{r|X\}}=\mathbbm{P}_n\int \gamma^T(r)\left \{W_r-\mathbbm{P}_n\left (\frac{J_rW_rX^T}{K_n\{r|X\}} \right )V_r^{-1}XK\{r|X\}\right\}
  \frac{dM_{\widetilde{C}}\{r|X\}}{K\{r|X\}}
  $$
Collecting terms and minimizing the variance of the resulting martingale term leads to the following optimal $\gamma (r)$: 
 \begin{align*}
 \gamma(r)=&\left (E \left [ \left\{W_r-D_rV_r^{-1}XK(r|X)\right \}\left\{W_r-D_rV_r^{-1}XK(r|X)\right \}^T J_r\frac{\lambda_{\widetilde C}(r|X)}{K(r|X)^2}\right ]\right )^{-1} \\
 &\times E \left [\left\{b_t(P_n,Z)-H^t_rV_r^{-1}XK(r|X)\right \}\left\{W_r-D_rV_r^{-1}XK(r|X)\right \}^T J_r\frac{\lambda_{\widetilde C}(r|X)}{K(r|X)^2}\right ]
 \end{align*}
 where 
 $$
 D_r=E\left (\frac{J_rW_rX^T}{K\{r|X\}}\right );\; H^t_r=E\left\{ b_t(P,Z)J_rX^T\right\}.
 $$
 
\newpage

\section{Web Appendix C}

We  focus here on the irreversible illness-death model.


\label{sec:sim_illnessdeath}

Data are sampled from the following data-generating process:
 $A \mid L \sim \text{Ber}(\text{expit}(-0.5 + \beta \cdot L))$ with $L \sim \text{Unif}(0,1)$;
    $T^* \mid  A, L \sim \text{Exp}(\lambda_{01} + \lambda_{0D})$ with $\lambda_{01} = 0.04 \cdot\exp(\gamma \cdot L + A)$  and $\lambda_{0D} = 0.02 \cdot\exp(\log(2) \cdot L + A)$;
    $\delta_1 \sim \text{Ber}\big(\frac{\lambda_{01}}{\lambda_{01} + \lambda_{0D}}\big)$; $T_D = T^* + \delta_1 \cdot U$ with $U \sim \text{Exp}(\lambda_{1D})$ and $\lambda_{1D} = 0.05 \cdot\exp(\gamma \cdot L + A)$;
     $T_1 = T^*$ if $\delta_1 = 1$; 
    $\widetilde{C} \mid  L \sim \text{Exp}(\lambda_{\widetilde{C}})$ with $\lambda_{\widetilde{C}} = \alpha \cdot\exp(A + \theta \cdot \mathbbm{I}(L > 0.5))$.
We recall that $\text{expit}(x) := \exp(x) / [1 + \exp(x)]$, and we set $\beta = 1$, $\gamma = \log(2)$, $\theta = 1$ and $\alpha = \{0.01, 0.03, 0.05\}$, which correspond approximately to a censoring proportion of about 27\%, 54\% and 67\%.
We then compute the fully efficient one-step estimator in (5) 
under different scenarios, the estimator with high efficiency in (7) 
and we compare the obtained results.

\subsection{The fully efficient one-step estimator}
\label{sec:sim_illnessdeath:onestep}
The propensity score is estimated through a logistic regression model, 
while the transition and censoring hazards are estimated using a Cox regression model. 
With the aim of showing the double robustness and asymptotic properties of the one-step estimator derived in (5), 
wherein (4) 
we employ (6) and $H_t(P;L)$ 
when fitting the working models we consider the following scenarios: 
    (i) All models 
    correctly specified ($\beta \neq 0$, $\gamma \neq 0$, $\theta \neq 0$);
    (ii) Propensity score misspecified ($\beta = 0$, $\gamma \neq 0$, $\theta \neq 0$);
    (iii) $\Lambda_{01}$ and $\Lambda_{1D}$ misspecified ($\beta \neq 0$, $\gamma = 0$,  $\theta \neq 0$);
    (iv) $\Lambda_{01}$, $\Lambda_{1D}$ and propensity score misspecified ($\beta = 0$,  $\gamma = 0$, $\theta \neq 0$);
    (v) 
    $\Lambda_{\widetilde{C}}$ misspecified ($\beta \neq 0$, $\gamma \neq 0$, $\theta = 0$);
    (vi) $\Lambda_{\widetilde{C}}$, $\Lambda_{01}$, $\Lambda_{1D}$ and propensity score misspecified ($\beta = 0$, $\gamma = 0$, $\theta = 0$).
For each scenario, we set the sample size to $1000$, the time horizon to $t = 10$ and we replicate the estimation procedure 1000 times. 

In Table \ref{tab:case_illnessdeath_censored_onestep} 
we report results for computed one-step estimator $\widehat \psi_t^{os}$, along with its building blocks, the plug-in estimator and the de-biasing term, across scenarios (i)-(vi) and different censoring hazards; results are shown for $g(\cdot) = \sqrt[3]{\cdot}$ because, among the simplest transformations, it effectively addresses the issue of skewed distributions caused by early deaths mentioned earlier.

\begin{table}
\centering 
\small
\begin{tabular}{@{}llcccccccc@{}}
\toprule 
 &  & & $\widehat \psi_t^{os}$ &  & & \textbf{Bias} & \textbf{SD} & \textbf{SE} & \textbf{Cov} \\
\cmidrule(l{0pt}r{2pt}){4-6}  
  &  & & & \textbf{Plug-in} & \textbf{De-bias} &  &  &  &  \\
  \midrule
\multirow{6}{*}{(i) All correct} 
& \multirow{2}{*}{$\alpha = 0.01$}
  & $A=1$  & 0.342 & 0.348 & -0.006 & 0.000 & 0.015 & 0.015 & 0.949\\
& & $A=0$  & 0.192 & 0.191 & 0.001 & 0.001 & 0.012 & 0.012 & 0.954 \\
\cmidrule{2-10}
& \multirow{2}{*}{$\alpha = 0.03$} 
  & $A=1$  & 0.341 & 0.348 & -0.007 & -0.002 & 0.018 & 0.019 & 0.956\\
& & $A=0$  & 0.191 & 0.188 & 0.003 & -0.001 & 0.015 & 0.015 & 0.963\\
\cmidrule{2-10}
& \multirow{2}{*}{$\alpha = 0.05$} 
  & $A=1$  & 0.337 & 0.346 & -0.009 & -0.005 & 0.033 & 0.027 & 0.952\\
& & $A=0$  & 0.188 & 0.183 & 0.005 & -0.004 & 0.019 & 0.019 & 0.951\\
\midrule
\multirow{6}{*}{(ii) Propen. score (PS) missp.} 
& \multirow{2}{*}{$\alpha = 0.01$} 
  & $A=1$  & 0.342 & 0.348 & -0.006 & -0.001 & 0.015 & 0.015 & 0.949\\
& & $A=0$  & 0.192 & 0.191 & 0.001 & 0.001 & 0.012 & 0.012 & 0.944\\
\cmidrule{2-10}
& \multirow{2}{*}{$\alpha = 0.03$} 
  & $A=1$  & 0.340 & 0.348 & -0.008 & -0.002 & 0.019 & 0.019 & 0.960\\
& & $A=0$  & 0.191 & 0.188 & 0.003 & 0.000 & 0.015 & 0.015 & 0.955\\
\cmidrule{2-10}
& \multirow{2}{*}{$\alpha = 0.05$} 
  & $A=1$  & 0.337 & 0.346 & -0.009 & -0.006 & 0.034 & 0.028 & 0.954\\
& & $A=0$  & 0.188 & 0.183 & 0.005 & -0.003 & 0.018 & 0.018 & 0.949\\
\midrule
\multirow{6}{*}{(iii) $\Lambda_{01}$ and $\Lambda_{1D}$ missp.} 
& \multirow{2}{*}{$\alpha = 0.01$} 
  & $A=1$  & 0.342 & 0.350 & -0.008 & 0.000 & 0.015 & 0.015 & 0.947\\
& & $A=0$  & 0.192 & 0.186 & 0.006 & 0.001 & 0.012 & 0.012 & 0.953\\
\cmidrule{2-10}
& \multirow{2}{*}{$\alpha = 0.03$} 
  & $A=1$  & 0.341 & 0.345 & -0.005 & -0.002 & 0.018 & 0.019 & 0.953\\
& & $A=0$  & 0.191 & 0.180 & 0.010 & -0.001 & 0.015 & 0.015 & 0.957\\
\cmidrule{2-10}
& \multirow{2}{*}{$\alpha = 0.05$} 
  & $A=1$  & 0.337 & 0.340 & -0.003 & -0.005 & 0.032 & 0.027 & 0.952\\
& & $A=0$  & 0.187 & 0.172 & 0.015 & -0.004 & 0.019 & 0.019 & 0.945\\
\midrule
\multirow{6}{*}{(iv) $\Lambda_{01}$, $\Lambda_{1D}$ and PS missp.} 
& \multirow{2}{*}{$\alpha = 0.01$} 
  & $A=1$  & 0.349 & 0.350 & -0.001 & 0.006 & 0.015 & 0.015 & 0.928\\
& & $A=0$  & 0.188 & 0.186 & 0.002 & -0.004 & 0.012 & 0.012 & 0.945\\
\cmidrule{2-10}
& \multirow{2}{*}{$\alpha = 0.03$} 
  & $A=1$  & 0.347 & 0.345 & 0.002 & 0.005 & 0.019 & 0.019 & 0.936\\
& & $A=0$  & 0.186 & 0.180 & 0.006 & -0.005 & 0.015 & 0.015 & 0.946\\
\cmidrule{2-10}
& \multirow{2}{*}{$\alpha = 0.05$} 
  & $A=1$  & 0.343 & 0.340 & 0.004 & 0.001 & 0.034 & 0.028 & 0.942\\
& & $A=0$  & 0.183 & 0.172 & 0.011 & -0.008 & 0.019 & 0.018 & 0.927\\
\midrule
\multirow{6}{*}{(v) $\Lambda_{\widetilde{C}}$ missp.} 
& \multirow{2}{*}{$\alpha = 0.01$} 
  & $A=1$  & 0.339 & 0.348 & -0.009 & -0.003 & 0.015 & 0.015 & 0.942\\
& & $A=0$  & 0.192 & 0.191 & 0.001 & 0.001 & 0.012 & 0.012 & 0.948\\
\cmidrule{2-10}
& \multirow{2}{*}{$\alpha = 0.03$} 
  & $A=1$  & 0.333 & 0.348 & -0.015 & -0.010 & 0.019 & 0.019 & 0.930\\
& & $A=0$  & 0.193 & 0.188 & 0.005 & 0.002 & 0.014 & 0.014 & 0.957 \\
\cmidrule{2-10}
& \multirow{2}{*}{$\alpha = 0.05$} 
  & $A=1$  & 0.325 & 0.346 & -0.021 & -0.018 & 0.027 & 0.026 & 0.921\\
& & $A=0$  & 0.194 & 0.183 & 0.011 & 0.002 & 0.018 & 0.017 & 0.957\\
\midrule
\multirow{6}{*}{(vi) $\Lambda_{\widetilde{C}}$, $\Lambda_{01}$, $\Lambda_{1D}$ and PS mis.} 
& \multirow{2}{*}{$\alpha = 0.01$} 
  & $A=1$  & 0.345 & 0.350 & -0.005 & 0.002 & 0.015 & 0.015 & 0.938\\
& & $A=0$  & 0.185 & 0.186 & -0.001 & -0.006 & 0.012 & 0.012 & 0.928\\
\cmidrule{2-10}
& \multirow{2}{*}{$\alpha = 0.03$} 
  & $A=1$  & 0.335 & 0.345 & -0.010 & -0.007 & 0.018 & 0.018 & 0.935\\
& & $A=0$  & 0.181 & 0.180 & 0.001 & -0.010 & 0.015 & 0.015 & 0.876 \\
\cmidrule{2-10}
& \multirow{2}{*}{$\alpha = 0.05$} 
  & $A=1$  & 0.324 & 0.340 & -0.016 & -0.018 & 0.025 & 0.024 & 0.900 \\
& & $A=0$  & 0.177 & 0.172 & 0.004 & -0.015 & 0.018 & 0.018 & 0.840\\
\bottomrule
\end{tabular}
\caption{Results for the \textit{one-step estimator} $\widehat \psi_t^{os}$ in the illness-death case at time point $t=10$, with $g(\cdot) = \sqrt[3]{\cdot}$ across scenarios (i)-(vi) and different censoring hazards. With respect to the one-step estimator $\widehat \psi_t^{os}$, we report its building blocks (plug-in and de-biasing terms), its bias with respect to the true value (Bias), its standard deviation (SD), its empirical standard error (SE), and its coverage at the 95\% confidence level (Cov). The sample size is set to 1000 and the estimation procedure is replicated 1000 times.}
\label{tab:case_illnessdeath_censored_onestep}
\end{table}

In scenarios (i) and (ii), both one-step and plug-in estimators are consistent, with coverage rates closely aligning with the nominal level. In scenario (iii), the one-step estimator remains consistent, while the plug-in estimator exhibits bias. 
Nonetheless, the coverage rate remains close to the nominal level, demonstrating alignment with the double robustness property. As expected, in scenario (iv), both estimators yield biased estimates.
For scenario (v), both one-step and plug-in estimators demonstrate consistency and results mirror those of scenario (ii).
Lastly, in scenario (vi), both one-step and plug-in estimators display bias. 
As a general trend across scenarios, higher censoring rates correspond to higher standard errors.

\subsection{The consistent estimator with high efficiency}

Focusing now only on scenario (i), in Table \ref{tab:case_illnessdeath_censored_estimator} we showcase results obtained by the estimator $\widehat \psi_t$ given in (7). 
Also in this case, we present results for $g(\cdot) = \sqrt[3]{\cdot}$ across different censoring hazards, we set the sample size to 1000, the time horizon to $t=10$ and replicate the estimation 1000 times.
We report the computed consistent estimator with high efficiency $ \widehat \psi_t$ presented in
(7), 
along with its component $ \widetilde \psi_t$ in (8) 
for comparison. The reported estimates are obtained using a model for the outcome that includes $A$, $L$, and their interaction term. Similar results were obtained when the interaction was omitted, thus they are not reported.
For the censoring model, a stratified Cox model based on $A$ and binary $L$ is employed.

\begin{table}
\centering 
\small
\begin{tabular}{@{}llcccccccccc@{}}
\toprule 
  & & &  \multicolumn{5}{c}{$\widehat \psi_t$ in (8) 
  } 
  & \multicolumn{3}{c}{$\widetilde \psi_t$ in (9) 
  }\\
\cmidrule(l{0pt}r{6pt}){4-8} \cmidrule(l{6pt}r{0pt}){9-11}
& & & \textbf{Mean} & \textbf{Bias} & \textbf{SD} & \textbf{SE} & \textbf{Cov} & \textbf{Mean} & \textbf{Bias} & \textbf{SD}\\ 
\midrule
\multirow{6}{*}{(i) All correct} 
& \multirow{2}{*}{$\alpha = 0.01$} 
  &   $A=1$  & 0.342 & 0.000 & 0.015 & 0.015 & 0.950 & 0.342 & -0.001 & 0.016\\
& & $A=0$  & 0.193 & 0.002 & 0.012 & 0.012 & 0.943 & 0.192 & 0.000 & 0.012\\
\cmidrule{2-11}
& \multirow{2}{*}{$\alpha = 0.03$} 
  &   $A=1$  & 0.343 & 0.001 & 0.020 & 0.019 & 0.946 & 0.342 & 0.000 & 0.021\\
& & $A=0$  & 0.197 & 0.005 & 0.014 & 0.014 & 0.933 & 0.193 & 0.001 & 0.014\\
\cmidrule{2-11}
& \multirow{2}{*}{$\alpha = 0.05$} 
  &   $A=1$  & 0.345 & 0.002 & 0.031 & 0.028 & 0.907 & 0.343 & 0.000 & 0.032\\
& & $A=0$  & 0.199 & 0.007 & 0.016 & 0.016 & 0.929 & 0.193 & 0.001 & 0.018 \\
\bottomrule
\end{tabular}
\caption{Results for the \textit{consistent estimator with high efficiency} $\widehat \psi_t$ in the illness-death case at time point $t=10$, with $g(\cdot) = \sqrt[3]{\cdot}$ for scenario (i) across different censoring hazards. With respect to $\widehat \psi_t$, we report its mean obtained across iterations (Mean), its bias with respect to the true value (Bias), its standard deviation (SD), its empirical standard error (SE), and its coverage at the 95\% confidence level (Cov).
For comparison, we report Mean, Bias and SD for its component $\widetilde \psi_t$. The sample size is set to 1000 and the estimation procedure is replicated 1000 times.}
\label{tab:case_illnessdeath_censored_estimator}
\end{table}

The results indicate that higher censoring rates correspond to higher standard errors and reduced coverage. Moreover, for the illness-death model, the improvement due to the censoring augmentation (transitioning from $\widetilde{\psi}_t$ to $\widehat{\psi}_t$) is small but noticeable, with a slightly lower SD for $\widehat{\psi}_t$. The benefit increases as the parameter $\alpha$ increases, as expected.
Furthermore, results obtained in Table \ref{tab:case_illnessdeath_censored_estimator} 
are very comparable to those in Table \ref{tab:case_illnessdeath_censored_onestep}.
We notice that standard errors are slightly lower when the fully efficient estimator is employed, as expected. The difference in this scenario is, however, negligible.

\newpage

\section{Web Appendix D}
\subsection*{Colorectal cancer study}
\label{Suppsec:case_study_clrc}
We employ the proposed estimator to analyze follow-up data for 150 metastatic colorectal cancer patients, randomly selected from the FFCD 2000-05 multicenter phase III clinical trial originally including 410 patients \citep{ducreux2011sequential}. Specifically, we examine the times of new lesion appearance, censored by terminal events (death or right-censoring).
Patients were randomized into two therapeutic strategies: combination (C) and sequential (S). Out of 150 patients, 73 (48.67\%) received the former, 77 (51.33\%) the latter.
The dataset
includes the baseline characteristics  
age ($<50$, $50$-$69$ or $>69$ years), WHO performance status 
(0, 1 or 2), and previous resection 
of the primary tumor (Yes or No).

\begin{figure}[H]%
\centering
\subfloat[Stratification by \texttt{treatment}]{\includegraphics[width = 0.48\columnwidth]{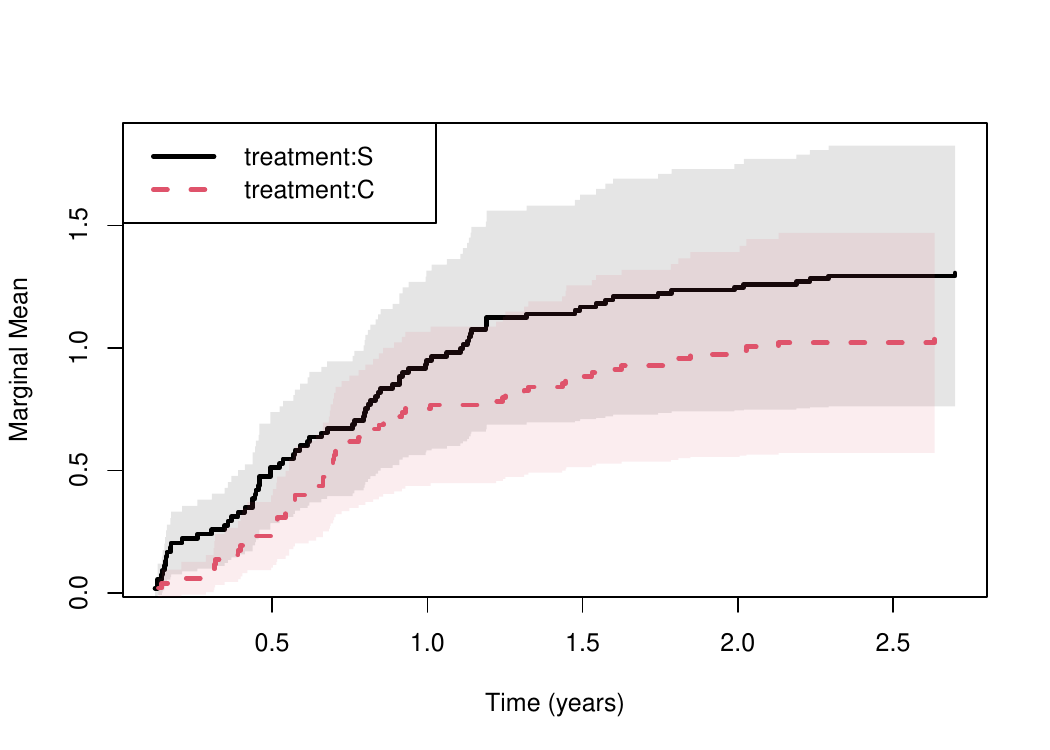}} \hspace{0.1cm}
\subfloat[Stratification by \texttt{prev.resection}]{\includegraphics[width = 0.48\columnwidth]{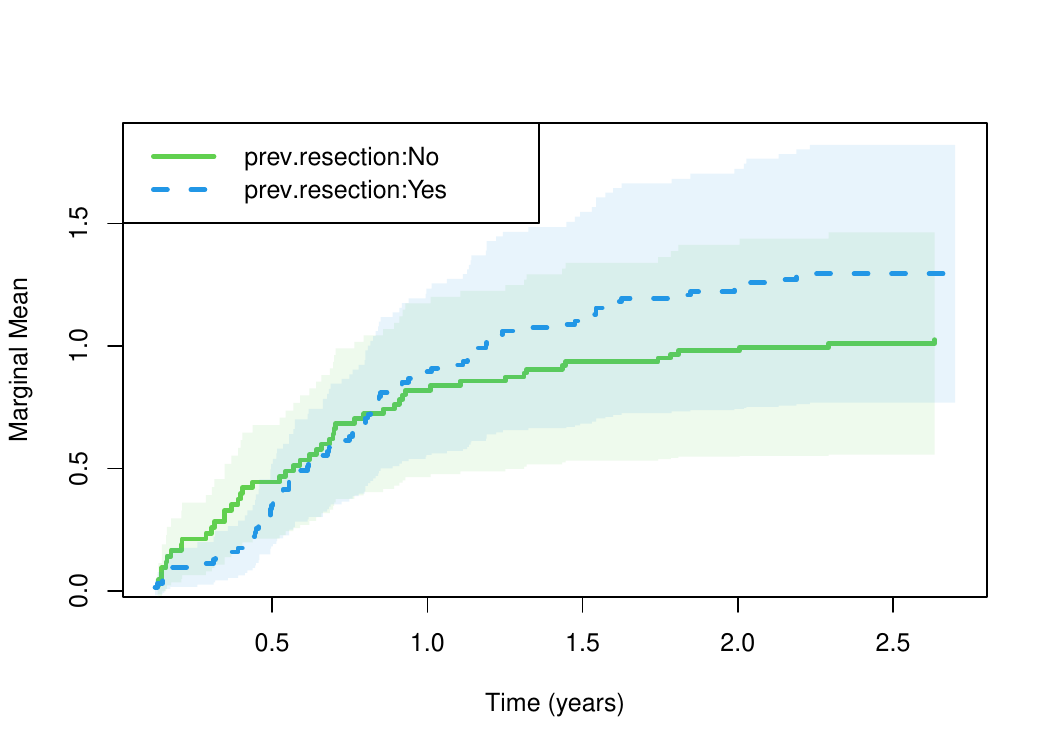}} \\
\subfloat[Stratification by \texttt{age}]{\includegraphics[width = 0.48\columnwidth]{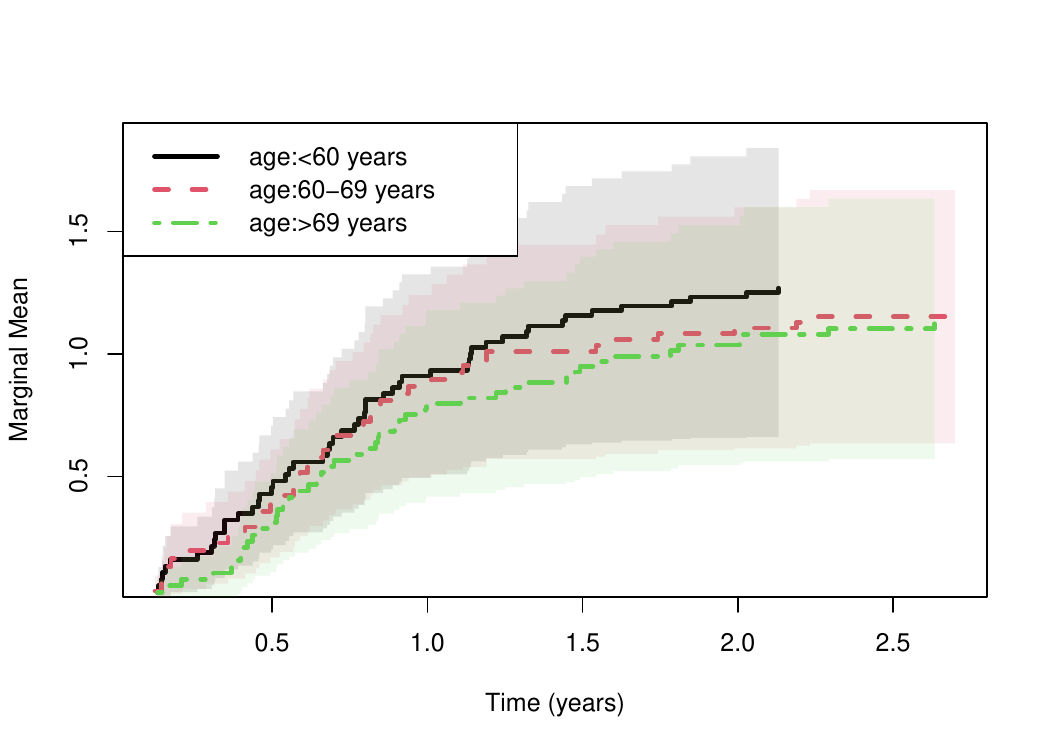}} \hspace{0.1cm}
\subfloat[Stratification by \texttt{who.PS}]{\includegraphics[width = 0.48\columnwidth]{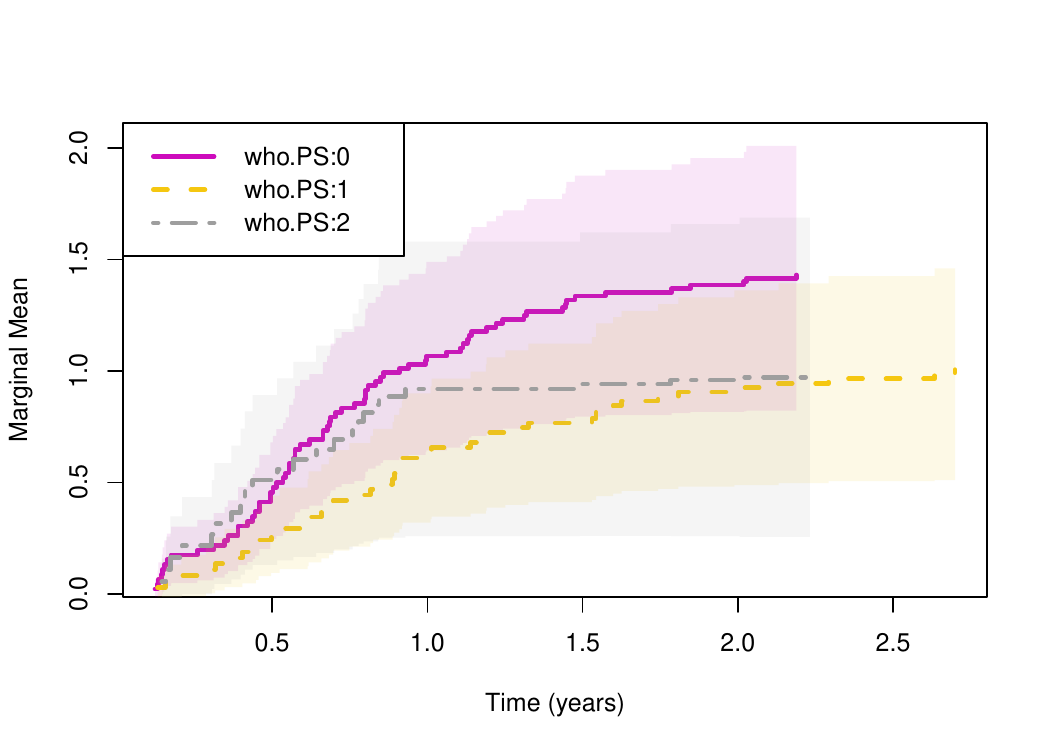}} \\
\caption{\textit{Colorectal cancer study}. Graphical visualization of the marginal mean of expected number of recurrent events (and its $95\%$ confidence interval), stratified by (i) treatment, 
(ii)  previous resection of primate tumor, 
(iii) age 
and (iv) WHO performance status.}
\label{fig:explorativeanalysis_casestudy}
\end{figure}

A graphical inspection
of the marginal mean of new lesion appearances over time, stratifying by various baseline characteristics, is possible in Figure \ref{fig:explorativeanalysis_casestudy}.

Over a median follow-up of 1.2 years, 64 patients (83.11\%) receiving treatment S died with an average of 1.03 new lesions per patient. 
In comparison, 57 patients (78.08\%) receiving treatment C died, with an average of 0.82 new lesions per patient. 
This preliminary analysis suggests that patients undergoing treatment S experience a slightly higher average number of new lesion appearances 
and have a higher mortality rate compared to those receiving treatment C.
However, when assessing the effect of treatment on new lesion appearances, it is crucial to account for the differential survival rates. Failing to do so could lead to incomplete or misleading conclusions. Notably, the consistent position of the black curve of group S above the dashed red line in Figure \ref{fig:explorativeanalysis_casestudy}, coupled with the higher mortality for S, suggests that patients in arm S may have less time to develop more lesions due to earlier death. This observation could indicate a potentially better outcome for treatment C. Further analysis is needed to properly adjust for these survival differences and ensure an accurate and significant assessment of treatment effects.

\begin{figure}
     \centering
     \begin{subfigure}[b]{0.48\textwidth}
         \centering
         \includegraphics[width=\textwidth]{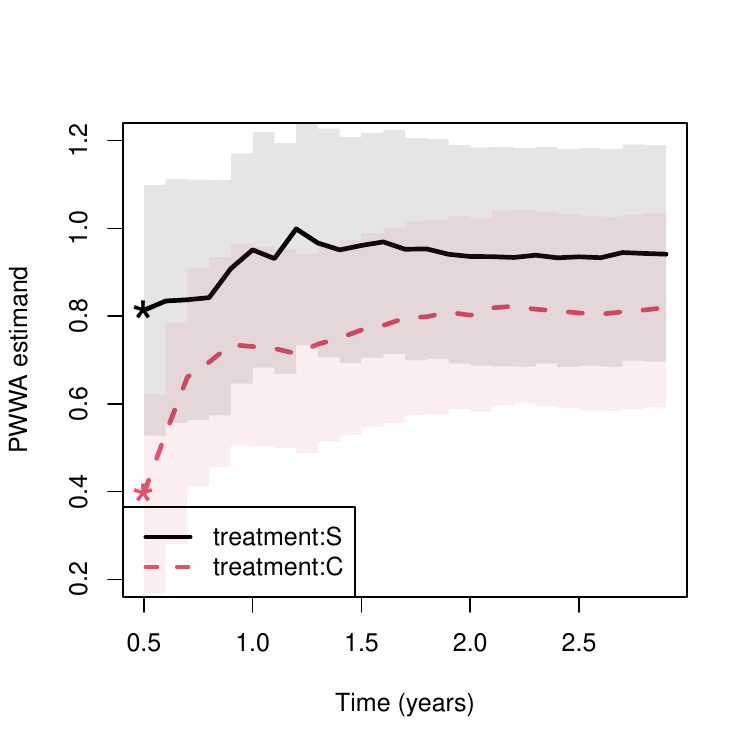}
         \caption{PWWA estimand.}
         \label{fig:colorectal-PWWA}
     \end{subfigure}
     \hfill
     \begin{subfigure}[b]{0.48\textwidth}
         \centering
         \includegraphics[width=\textwidth]{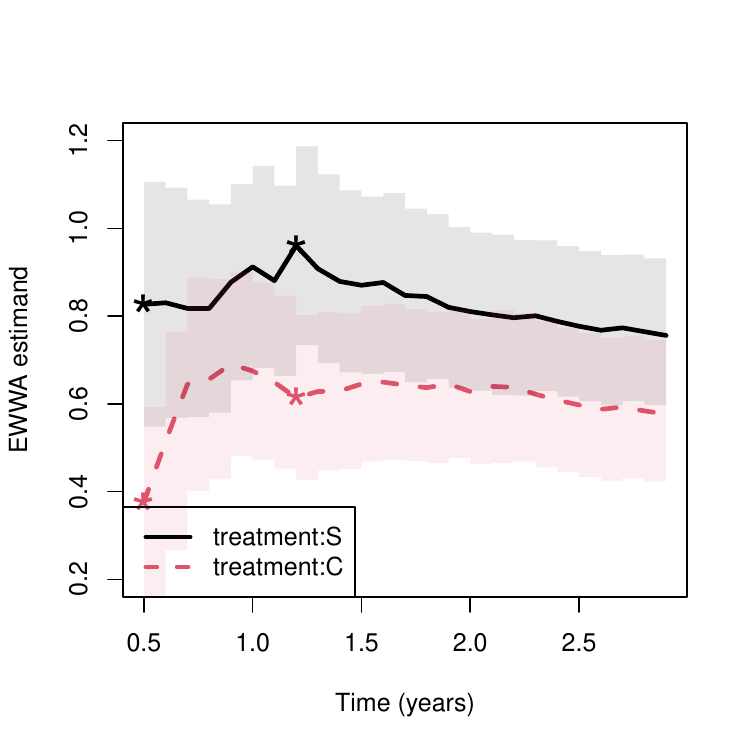}
         \caption{EWWA estimand.}
         \label{fig:colorectal-EWWA}
     \end{subfigure}
     \vspace{0.25cm}
     \caption{\textit{Colorectal cancer study}. Graphical visualization of PWWA and EWWA estimands (and their $95\%$ confidence intervals) over 5 semesters. Asterisks indicate a p-value $< 0.05$ for the test with null hypothesis \lq\lq no difference among S and C\rq\rq.}
     \label{fig:colorectal}
\end{figure}

Thus, we employ the proposed estimator 
to estimate the effect of the treatment 
on the average number of new lesions appearances before the terminal event over the time window $[0,t]$ years, with $t=0.5,\dots,2.9$. 
The results at various time points are displayed in Figure \ref{fig:colorectal-PWWA}, along with their respective 95\% confidence intervals, being $g(\cdot)$ chosen as the identity. Asterisks indicate the time points at which the two therapeutic strategies show statistically significant differences at a significance level of $0.05$.
The reported estimates 
are derived using both a model for the outcome and a Cox model for censoring that include treatment, 
baseline covariates, and their interaction terms.

\nothere{
Results are reported in Web Table \ref{tab:case_studyCC}, for two different choices of the transformation function $g(\cdot)$: identity and cube root. 
As noted earlier, the cube root is one of the simplest transformations that helps address skewness that may be introduced by early deaths, ensuring a more balanced evaluation across patients with varying survival times. However, this transformation significantly affects the ease of interpretation of the results derived from the while-alive estimand.
On the other hand, the identity function improves interpretability and facilitates direct comparison with Web Figure \ref{fig:explorativeanalysis_casestudy}, and for this reason, it will be the object of our focus. 
}


\nothere{
\begin{table}
\centering 
\small
\begin{tabular}{@{}llcccccc@{}}
\toprule 
& & \multicolumn{2}{c}{$t=0.5$} & \multicolumn{2}{c}{$t=1.5$} &  \multicolumn{2}{c}{$t=2.5$} \\ 
\cmidrule(l{3pt}r{3pt}){3-4} \cmidrule(l{3pt}r{3pt}){5-6} \cmidrule(l{3pt}r{3pt}){7-8} 
& & Est (SE) &  $p$-value & Est (SE) & $p$-value & Est (SE) & $p$-value \\ 
\midrule
\multirow{3}{*}{$g(\cdot) = \cdot$} 
  &   \texttt{treatment}:S  & 0.813 (0.146) & 0 & 0.961 (0.131) & 0 &  0.935 (0.126) & 0 \\
& \texttt{treatment}:C  & 0.396 (0.116) & 0 & 0.768 (0.113) & 0 & 0.807 (0.113) & 0 \\
\cmidrule{2-8}
& Difference (S - C) & 0.418 (0.188) & 0.026 & 0.193 (0.176) &  0.272  & 0.128 (0.173) & 0.458 \\
\midrule
\multirow{3}{*}{$g(\cdot) = \sqrt[3]{\cdot}$} 
  &   \texttt{treatment}:S  & 4.006 (0.124) & 0 & 3.977 (0.125) & 0 & 3.903 (0.131) & 0 \\
& \texttt{treatment}:C  & 3.933 (0.125) & 0 & 3.918 (0.125) & 0 &  3.892 (0.128) & 0 \\
\cmidrule{2-8}
& Difference (S - C) & 0.073 (0.177) & 0.680 & 0.059 (0.177) & 0.739 &  0.011 (0.191) &  0.953 \\
\bottomrule
\end{tabular}
\caption{\textit{Colorectal cancer case study} results for the PWWA estimand, for two different choices of $g(\cdot)$. Est, estimate of the PWWA estimand; SE, standard error.
}
\label{tab:case_studyCC}
\end{table}
}

Results for $t>0.5$ 
suggest that there is not sufficient evidence to reject the null hypothesis, which posits no difference between the two therapeutic strategies, leading to the conclusion that treatment strategies do not have a statistically significant different impact on the expected value for the number of new lesion appearances over the time-alive up to $t$ years.
Only at $t=0.5$ there is enough evidence at 0.05 to reject the null, 
allowing to conclude that treatment strategies have a statistically significant different impact, and treatment C should be preferred.

As shown in Figure \ref{fig:colorectal-EWWA}, the analyzes of the colorectal cancer study based on the EWWA estimand yield overall consistent conclusions to those for the PWWA estimand.

\newpage

\newpage

\section*{Web Table 3}
\subsection*{}

\begin{table}[H]
\centering 
\small
\begin{tabular}{@{}llcccccccccc@{}}
\toprule 
& & & & \multicolumn{4}{c}{PWWA} & \multicolumn{4}{c}{EWWA} \\ 
\cmidrule(l{3pt}r{3pt}){5-8} \cmidrule(l{3pt}r{3pt}){9-12}
& $v$ & $\theta$ & $s_D$  & \textbf{Mean} & \textbf{SD} & \textbf{SE}  & \textbf{Power} & \textbf{Mean} & \textbf{SD} & \textbf{SE} & \textbf{Power}\\ 
\midrule
\multirow{8}{*}{(a)} & \multirow{4}{*}{$1$} 
& \multirow{2}{*}{1} 
   &  1 & -0.000 & 0.032 & 0.032 & 0.045 & -0.001 & 0.074 & 0.074 & 0.051 \\ 
& & & 4 & -0.000 & 0.035 & 0.035 & 0.042 & 0.000 & 0.065 & 0.065 & 0.052 \\ 
\cmidrule{3-12}
& & \multirow{2}{*}{2} 
  & 1   & -0.000 & 0.034 & 0.034 & 0.050 & 0.000 & 0.086 & 0.086 & 0.051 \\ 
& & & 4 & -0.000 & 0.036 & 0.035 & 0.056 & 0.001 & 0.064 & 0.064 & 0.050 \\ 
\cmidrule{2-12}
& \multirow{4}{*}{$0$} 
& \multirow{2}{*}{1} 
  & 1   & -0.000 & 0.031 & 0.032 & 0.045 & -0.000 & 0.084 & 0.083 & 0.052 \\  
& & & 4 & -0.001 & 0.034 & 0.034 & 0.048 & -0.001 & 0.091 & 0.090 & 0.049 \\  
\cmidrule{3-12}
& & \multirow{2}{*}{2} 
  & 1   & -0.000 & 0.034 & 0.034 & 0.051 & 0.001 & 0.107 & 0.108 & 0.048 \\  
& & & 4 & -0.000 & 0.034 & 0.034 & 0.050 & 0.002 & 0.115 & 0.116 & 0.047 \\  
\midrule
\multirow{8}{*}{(b)} & \multirow{4}{*}{$1$} 
& \multirow{2}{*}{1} 
   &  1 & -0.000 & 0.033 & 0.033 & 0.049 & -0.001 & 0.072 & 0.072 & 0.054 \\ 
& & & 4 &  -0.000 & 0.035 & 0.036 & 0.048 & 0.000 & 0.059 & 0.060 & 0.049 \\ 
\cmidrule{3-12}
& & \multirow{2}{*}{2} 
  & 1   & -0.000 & 0.034 & 0.034 & 0.052 & 0.001 & 0.082 & 0.082 & 0.048 \\  
& & & 4 & 0.000 & 0.036 & 0.035 & 0.053 & 0.000 & 0.058 & 0.058 & 0.047 \\  
\cmidrule{2-12}
& \multirow{4}{*}{$0$} 
& \multirow{2}{*}{1} 
  & 1   & -0.000 & 0.032 & 0.032 & 0.050 & -0.000 & 0.082 & 0.082 & 0.052 \\ 
& & & 4 & -0.000 & 0.034 & 0.034 & 0.048 & -0.001 & 0.088 & 0.087 & 0.050 \\ 
\cmidrule{3-12}
& & \multirow{2}{*}{2} 
  & 1   & -0.000 & 0.034 & 0.034 & 0.049 & 0.002 & 0.106 & 0.106 & 0.047 \\
& & & 4 & -0.000 & 0.034 & 0.034 & 0.047 & 0.002 & 0.111 & 0.112 & 0.050 \\
\midrule
\multirow{8}{*}{(c)} & \multirow{4}{*}{$1$} 
& \multirow{2}{*}{1} 
   &  1 & 0.000 & 0.031 & 0.032 & 0.045 & 0.001 & 0.098 & 0.098 & 0.051 \\ 
& & & 4 & 0.000 & 0.035 & 0.035 & 0.050 & 0.000 & 0.109 & 0.108 & 0.051 \\ 
\cmidrule{3-12}
& & \multirow{2}{*}{2} 
  & 1   & 0.001 & 0.035 & 0.035 & 0.049 & 0.003 & 0.120 & 0.121 & 0.044 \\ 
& & & 4 & 0.001 & 0.037 & 0.037 & 0.053 & 0.004 & 0.115 & 0.117 & 0.047 \\ 
\cmidrule{2-12}
& \multirow{4}{*}{$0$} 
& \multirow{2}{*}{1} 
  & 1   & -0.000 & 0.032 & 0.032 & 0.048 & 0.001 & 0.101 & 0.101 & 0.051 \\ 
& & & 4 & 0.001 & 0.035 & 0.036 & 0.047 & 0.001 & 0.128 & 0.126 & 0.054 \\ 
\cmidrule{3-12}
& & \multirow{2}{*}{2} 
  & 1   & 0.001 & 0.036 & 0.036 & 0.050 & 0.004 & 0.132 & 0.133 & 0.045 \\  
& & & 4 & 0.000 & 0.038 & 0.038 & 0.049 & 0.003 & 0.162 & 0.164 & 0.049 \\  
\bottomrule
\end{tabular}
\caption{Results comparing PWWA and EWWA estimands across different simulation settings (a–b-c), dependence structures, and scaling factors, setting $\beta_1 = \beta_d = 0$. For each estimand, we report the average estimated causal effect for the contrast (0-1) (Mean), the standard deviation (SD), the empirical standard error (SE) and observed power, 
computed testing the null hypothesis of no causal contrast between groups, using a significance level of 0.05.
The sample size is set to 1000 and the estimation procedure is replicated 5000 times. 
}
\label{tab:simulationstudyTab2NULL}
\end{table}


\newpage

\bibliographystyle{chicago} 
\bibliography{bibliography}